\documentclass[twocolumn,english]{autart}    

\usepackage{graphicx}          
\usepackage{graphics}                               
\usepackage{color}
\usepackage[dvips]{epsfig}    
\usepackage{amsfonts}       
\usepackage{amssymb}
\usepackage{amsmath}
\usepackage{amssymb}
\usepackage{wrapfig}
\usepackage{bm}
\usepackage{subfig}
\usepackage{threeparttable}
\usepackage[wide]{sidecap}
\usepackage{array}
\usepackage{algorithm}
\usepackage{algpseudocode}
\usepackage{lipsum,babel}
\usepackage{color}
\usepackage{setspace}
\usepackage{hyperref}
\newtheorem{thrm}{Theorem}
\newtheorem{lemm}{Lemma}
\newtheorem{deff}{Definition}

\newtheorem{assump}{Assumption}
\newtheorem{corol}{Corollary}

\newtheorem{remark}{Remark}

\newtheorem{proof}{Proof}
\newtheorem{problem}{Problem}

\renewcommand{\hat}[1]{\widehat{#1}}
\renewcommand{\tilde}[1]{\widetilde{#1}}

\def \exp {\mathrm{exp}}

\def \vec {\mathrm{vec}}

\def \R {\mathbb{R}}
\def \C {\mathbb{C}}
\def \tailconst {\widetilde{\beta}}
\def \tailcoeff {\beta}
\def \tailexp {\alpha}

\def \ssconstant {C}

\def \policy {\boldsymbol{\pi}}
\def \adaptivepolicy {\hat{\boldsymbol{\pi}}}
\def \optimalpolicy {\boldsymbol{\pi^\star}}
\def \para {\boldsymbol{\theta}}
\def \rrate {\gamma}

\def \causal {\boldsymbol{\mathcal{C}}}
\def \adaptive {\boldsymbol{\mathcal{A}}}

\newcommand{\Mnorm}[2]{{\left\vert\kern-0.35ex\left\vert\kern-0.35ex\left\vert #1 
		\right\vert\kern-0.35ex\right\vert\kern-0.35ex\right\vert}}
\newcommand{\Opnorm}[3]{{\left\vert\kern-0.35ex\left\vert\kern-0.35ex\left\vert #1 
		\right\vert\kern-0.35ex\right\vert\kern-0.35ex\right\vert}_{#2 \to #3}}
\newcommand{\norm}[2]{{\left\vert\kern-0.35ex\left\vert #1 
		\right\vert\kern-0.35ex\right\vert}}
\newcommand{\eigmax}[1]{\left| \lambda_{\max} \left( #1 \right)\right|}
\newcommand{\eigmin}[1]{\left| \lambda_{\min} \left( #1 \right)\right|}
\newcommand{\tr}[1]{\mathrm{tr} \left( #1 \right)}
\newcommand{\PP}[1]{\mathbb{P} \left(#1\right)}
\newcommand{\E}[1]{\mathbb{E} \left[#1\right]}

\newcommand{\innerproductminconstant}[1]{\psi_0}

\newcommand{\regret}[2]{\mathcal{R}_{#1} \left(#2\right)}

\newcommand{\tempparaspace}[1]{\Gamma_{#1}}
\newcommand{\Kmatrix}[1]{K\left(#1\right)}
\newcommand{\Lmatrix}[1]{L\left(#1\right)}
\newcommand{\extendedLmatrix}[1]{\widetilde{L}\left(#1\right)}

\newcommand{\instantcost}[2]{c_{#1}\left(#2\right)}

\newcommand{\optpara}[1]{\boldsymbol{\tilde{\theta}}_{#1}}

\newcommand{\estpara}[1]{\hat{\boldsymbol{\theta}}_{#1}}

\newcommand{\nullspace}[1]{\mathcal{N}_{#1}}

\newcommand{\riccatiOp}[2]{\Psi_{#1}\left(#2\right)}
\newcommand{\covmat}[1]{\Sigma_{#1}}
\newcommand{\empiricalcovmat}[1]{\widehat{\Sigma}_{#1}}
\newcommand{\ubound}[1]{\overline{#1}}
\newcommand{\lbound}[1]{\underline{#1}}
\newcommand{\totalbound}[1]{{\mathcal{E}_{#1}}}
\newcommand{\sequence}[2]{\tau_{#1}\left(#2\right)}
\newcommand{\optstate}[1]{a\left(#1\right)}
\newcommand{\surnoise}[1]{b\left(#1\right)}

\newif\ifarxiv
\arxivtrue

\ifarxiv

\else

\fi

\begin{document}

\begin{frontmatter}
	
	\title{Input Perturbations for Adaptive Control and Learning} 
	
	\ifarxiv 
	\author[]{Mohamad Kazem Shirani Faradonbeh},   
	\author[]{Ambuj Tewari},
	\author[]{George Michailidis} 
	\fi             

	
	\begin{abstract}
	{This paper studies adaptive algorithms for simultaneous regulation (i.e., control) and estimation (i.e., learning) of Multiple Input Multiple Output (MIMO) linear dynamical systems. It proposes \emph{practical}, easy to implement control policies based on \emph{perturbations} of input signals. Such policies are shown to achieve a worst-case regret that scales as the square-root of the time horizon, and holds uniformly over time. Further, it discusses specific settings where such greedy policies attain the information theoretic lower bound of logarithmic regret. To establish the results, recent advances on self-normalized martingales together with a novel method of policy decomposition are leveraged.}
	\end{abstract}
\end{frontmatter}



\section{Introduction}
MIMO systems with linear dynamics represent canonical models in system engineering \cite{kailath1980linear}. Their linear structure renders them amenable to rigorous mathematical analysis of their performance, as well as computationally feasible implementation of control policies~\cite{meyn2008control,faradonbeh2016optimality}. Further, they provide insights on how to deal with nonlinear dynamical models \cite{marino1995nonlinear,li2013stabilization}, and accurately represent the behavior of nonlinear systems around their operating equilibrium \cite{lazic2018data}. In addition, they are also providing benchmarks for prediction tasks~\cite{pesaran2005small,faradonbeh2018finite}. 

The standard formulation is that the system is characterized by autoregressive dynamics (in discrete time), while its operating cost is described by a quadratic function~\cite{bertsekas1995dynamic}. An extensive and mature literature exists on optimal Linear-Quadratic (LQ) policies when the dynamical model is exactly known~\cite{dorato1995linear}. However, for systems whose dynamics are unknown, an adaptive policy is needed in order to simultaneously learn (identify) the model parameters, and control (regulate) its behavior. This gives rise to designing adaptive Linear-Quadratic-Regulators (LQR) in order to balance accurate learning (exploration) with efficient regulation (exploitation)~\cite{guo1991aastrom}. {A common approach in adaptive control uses the principle of optimism in the face of uncertainty by designing polices assuming that optimistic approximations of the unknown parameters correspond to the truth~\cite{lai1985asymptotically}. However, existing asymptotic \cite{campi1998adaptive,bittanti2006adaptive} and non-asymptotic \cite{abbasi2011regret,ibrahimi2012efficient,faradonbeh2017finite} analyses of optimism-based adaptive regulators indicate that the proposed algorithms are not computationally tractable. That is because non-convex matrix optimization problems, requiring high precision accuracy of the optimal solution, need to be solved repeatedly. In addition, optimism-based policies need to have access to information regarding spectral properties of the system's closed-loop matrix, as well as the statistics of the noise process.}

{On the other hand, a greedy policy (also referred to as Certainty Equivalent (CE)) uses the optimal feedback gain of the current estimates of the dynamics parameters. Then, upon collecting new input-output observations, this greedy policy learns the unknown parameters through a least squares procedure, and uses the above learning-planning steps in an alternating manner. Hence, CE based policies do not require the aforementioned additional knowledge that optimism-based adaptive policies demand. A recent result establishes that if one randomizes the greedy adaptive regulator, it provides a regret of asymptotically square-root magnitude (with respect to time). In fact, it suffices to obtain the least squares estimate (LSE) of the unknown dynamics matrix, and randomize the estimate by adding a statistically independent random matrix to it~\cite{faradonbeh2018optimality}. } 

{However, the \emph{non-asymptotic} performance of the above randomized greedy algorithm is not satisfactory, due to the presence of ({frequent} and {large}) fluctuations in the trajectory of the system. This renders the finite-time performance of this regulator sub-optimal, since the regret can become large over a short time horizon. Technically, the probability of ``the regret exceeding a factor of the optimal value" decays polynomially (thus, slowly) as a function of the magnitude of that factor. Hence, in order to ensure that with high probability the regret remains within a margin of the optimal rate, the size of the posited margin needs to be large. In conclusion, the magnitude of the regret can exceed the square root of time\footnote{For more details, see Theorem \ref{GreedyRates} and the ensuing discussion.}. Note that the above fluctuations do not compromise the asymptotic performance of randomized greedy regulators, as the time horizon grows to infinity.}

{This study presents the \emph{first finite-time} theoretical guarantee for such practical algorithms for both system identification and adaptive control. It establishes that a greedy algorithm, subject to a suitable input {\em perturbation}, achieves {finite-time} efficiency of cost minimization, as well as learning accuracy. Namely, it is shown that at \emph{all} time steps, the \emph{worst-case} regret scales at a square-root of time rate; in addition, the \emph{uniform} non-asymptotic learning rates of the unknown dynamics are provided. Further, in Section~\ref{general theory}, a comprehensive analysis of input perturbation, applicable to various problems in stochastic control~\cite{imer2006optimal,ouyang2018optimal} is presented. To obtain the results, tools from the theory of martingale concentration~\cite{abbasi2011improved,tropp2012user} are employed, together with an extension of the policy decomposition method~\cite{faradonbeh2018optimality} to general control laws with perturbed input.} 

The remainder of the paper is organized as follows. In Section \ref{problem statement}, we provide a precise formulation of the problem under consideration, and discuss the necessary preliminaries. We then address the behavior of the non-asymptotic regret for general control policies with perturbed inputs in Subsection~\ref{exploitsubsection}. Subsequently, the effect of the perturbation signal on the high probability estimation error of the true model is given in Subsection~\ref{exploresubsection}. In Section \ref{CE section}, we study the growth rate of the regret, as well as the accuracy of learning procedure for the perturbed greedy regulator. We also discuss the case of restricted uncertainty in Subsection~\ref{CE with side}, where side information (such as the support or the rank of the dynamics matrices) is available to the system's operator. 
\begin{remark}
	Unless otherwise explicitly stated, all stochastic statements in this work hold almost surely.
\end{remark}
The following {notation} will be used throughout this paper. For a matrix $A \in \C^{p \times q}$, $A'$ denotes its transpose. When $p=q$, the smallest (respectively largest) eigenvalue of $A$ (in magnitude) is denoted by $\lambda_{\min} (A)$ (respectively $\lambda_{\max}(A)$). For $v \in \C^d$, define the norm $\norm{v}{} = \left( \sum\limits_{i=1}^{d} \left| v_i \right|^2 \right)^{1/2}$. We also use the following notation for the operator norm of matrices. For $A \in \C^{p \times q}$ let, $\Mnorm{A}{2} = \sup \limits_{\norm{v}{2}=1} \norm{Av}{2}$. 
The symbol $\vee$ ($\wedge$) will be used to denote the maximum (minimum) of two or more quantities.

\section{Problem Formulation} \label{problem statement}
{The input-output evolution of the system is governed by the following stochastic dynamical model. Starting from an arbitrary initial $x(0) \in \R^{p}$, at time $t=0,1,\cdots$, the transition to the next state (or output) is according to
\begin{eqnarray}
x(t+1) &=& A_0x(t)+B_0u(t)+ w(t+1). \label{systemeq1}
\end{eqnarray}
Namely, $x(t+1)$ is determined by the current state vector $x(t) \in \R^{p}$, the input signal $u(t) \in \R^r$, and the noise component $w(t+1) \in \R^p$, the latter being a mean-zero random vector: $\E{w(t+1)}=0$. This study aims to address the problems of system identification and adaptive regulation, when the dynamics parameters $A_0,B_0$ are \emph{unknown}. Specifically, it aims to design the input signal to minimize the cost, and to learn the matrices $A_0,B_0$ accurately. We start by presenting some necessary background material, followed by a rigorous formulation of the problems under consideration.}

The transition matrix $A_0 \in \mathbb{R}^{p \times p}$ models the effect of the current state signal, while $B_0 \in \mathbb{R}^{p \times r}$ is the input matrix indicating the influence of the control signal. The noise vectors $\left\{ w(t) \right\}_{t=1}^\infty$ are assumed to be independent, but are not required to be identically distributed (strict-sense stationary), and can even be heteroscedastic (wide-sense non-stationary). Note that the independence assumption on the noise vectors is not restrictive and can be replaced by assuming that they form a martingale difference sequence, without impacting the results established in this work.

To proceed, let $Q \in \R^{p \times p}$ and $R \in \R^{r \times r}$ be positive definite symmetric matrices. Then, the instantaneous cost of control law $\policy$ at time $t$ is defined as
\begin{eqnarray}
\instantcost{t}{\policy}&=& x(t)'Qx(t) + u(t)'Ru(t). \label{systemeq2}
\end{eqnarray}
Intuitively speaking, a desired control policy minimizes the long-run average cost. For a precise definition, we first specify the families of causal and adaptive policies, denoted by $\causal$ and $\adaptive$, respectively. A \emph{causal} policy $\policy \in \causal$ determines the input based on the dynamics matrices $A_0,B_0$, the matrices $Q,R$ which are assessing the costs of the state and control signals, and the record of the both signals thus far. That is,
\begin{eqnarray*}
	u(t) = \policy \left( A_0, B_0 , Q, R , \left\{ x(i) \right\}_{i=0}^t, \left\{ u(j) \right\}_{j=0}^{t-1} \right),
\end{eqnarray*}
for all $t \geq 0$. Note that the above mapping can be a stochastic one. 

An {\em adaptive} operator does not have access to the dynamical parameters $A_0, B_0$. Therefore, $\adaptivepolicy \in \adaptive$ designs the input signal according to other available information:
\begin{eqnarray*}
	u(t) = \adaptivepolicy \left( Q, R , \left\{ x(i) \right\}_{i=0}^t, \left\{ u(j) \right\}_{j=0}^{t-1} \right).
\end{eqnarray*}
Obviously, it holds that $\adaptive \subsetneqq \causal$. The control objective is to minimize the {average} cost of the system, subject to the stochastic dynamics equation \eqref{systemeq1}. Naturally, all $\policy \in \causal$ are expected to target the same objective being determined by the context of the problem. Therefore, the cost matrices $Q,R$ are assumed known for design of all $\adaptivepolicy \in \adaptive$. The best performance among causal control laws belongs to $\optimalpolicy$, which attains the smallest average cost:
\begin{eqnarray*}
	\limsup\limits_{n \to \infty} n^{-1} \sum \limits_{t=0}^{n-1} {\instantcost{t}{\optimalpolicy}} =  \inf\limits_{\policy \in \causal} \limsup \limits_{n \to \infty} n^{-1} \sum \limits_{t=0}^{n-1} {\instantcost{t}{\policy}}.
\end{eqnarray*}
In order to find $\optimalpolicy$, one needs to solve a Riccati equation~\cite{faradonbeh2017finite}. To do so, we assume that the following necessary condition of stabilizability is satisfied. 
\begin{assump}
	There exists a feedback gain matrix $L \in \R ^{r \times p}$ such that $\eigmax{A_0+B_0L} < 1$. 
\end{assump}
\begin{remark}
	Henceforth, we employ the shorthand $\para \in \R^{p \times q}$ (where $q=p+r$), to denote the pair $\left[A,B\right]$ of dynamics matrices $A \in \R^{p \times p}$, $B \in \R^{p \times r}$. 
\end{remark}
Then, for the dynamics parameter $\para$, define the Riccati operator $\riccatiOp{\para}{\cdot}: \R^{p \times p} \to \R^{p \times p}$:
\begin{eqnarray*}
	\riccatiOp{\para}{K} = Q + A'KA - A' K B \left(B'KB+R\right)^{-1} B'KA.
\end{eqnarray*}
Suppose that $\Kmatrix{\para}$ solves the algebraic Riccati equation
\begin{eqnarray} \label{ricatti2}
\Kmatrix{\para}=\riccatiOp{\para}{\Kmatrix{\para}}.
\end{eqnarray}
It has been established that whenever $\para$ is stabilizable, the positive definite $\Kmatrix{\para}$ exists, and is unique~\cite{faradonbeh2018stabilization}. In addition, the recursive Riccati equation $K_{t+1}=\riccatiOp{\para}{K_t}$, is known to provide a sequence of matrices converging exponentially fast to $\Kmatrix{\para}$~\cite{faradonbeh2018stabilization}. Using the matrix $\Kmatrix{\para}$ in \eqref{ricatti2}, let
\begin{eqnarray} \label{ricatti1}
\Lmatrix{\para} = -\left(B'\Kmatrix{\para}B+R\right)^{-1} B'\Kmatrix{\para}A.
\end{eqnarray}
The linear time invariant feedback gain $\Lmatrix{\para_0}$ provides an optimal causal regulator for a system of dynamics parameter $\para_0$ \cite{faradonbeh2017finite}. It is also well-known that $\Lmatrix{\para_0}$ is a stabilizer; i.e. $\eigmax{A_0+B_0\Lmatrix{\para_0}}<1$. Henceforth, let $\optimalpolicy$ be 
\begin{eqnarray} \label{optpolicydeff}
\optimalpolicy: \:\:\: u(t)= \Lmatrix{\para_0} x(t), \:\:\: t=0,1,2,\cdots.
\end{eqnarray}
Optimality of $\optimalpolicy$ motivates us to consider adaptive regulators of the form $u(t)=L_t x(t)$. Note that since information on the unknown dynamics parameters is acquired over time, such adaptive regulators are time-varying. Further, to address identification of the unknown parameters, $\adaptivepolicy \in \adaptive$ is subject to exogenous randomness through additional perturbations. So, in general we study causal policies of the form 
\begin{eqnarray}
\policy : \:\:\: u(t)=L_t x(t)+ v(t), \:\:\: t=0,1,2,\cdots,
\end{eqnarray}
where $L_t$ is a $r \times p$ feedback gain matrix (determined according to $A_0, B_0, Q, R $, $\left\{ x(i) \right\}_{i=0}^t, \left\{ u(j) \right\}_{j=0}^{t-1}$), and the $r$ dimensional perturbation vectors are $\left\{ v(t) \right\}_{t=0}^\infty$. We denote the above policy by $\policy= \left\{ L_t ,v(t) \right\}_{t=0}^\infty \in \causal$. Similarly, the adaptive regulator $\adaptivepolicy= \left\{ \widehat{L}_t, v(t) \right\}_{t=0}^\infty \in \adaptive$ is defined according to the perturbation signal $v(t)$, together with the matrices $\left\{\widehat{L}_t\right\}_{t=0}^\infty$, where $\widehat{L}_t $ is a function of $Q, R ,  \left\{ x(i) \right\}_{i=0}^t, \left\{ u(j) \right\}_{j=0}^{t-1}$. 
\begin{remark} \label{purturbproperties}
	In this work, we study mean zero and statistically independent perturbation signals $\left\{ v(t) \right\}_{t=0}^\infty$, which also are independent of the noise process $\left\{ w(t) \right\}_{t=1}^\infty$.
\end{remark}

The efficiency of $\policy \in \causal$ is determined by comparing the accumulative cost of $\policy$ to that of $\optimalpolicy$ in~\eqref{optpolicydeff}. In fact, we investigate the following natural definition of {\em regret}: 
\begin{eqnarray} \label{regretdeff}
\regret{n}{\policy} = \sum \limits_{t=0}^{n-1} \left[ \instantcost{t}{\policy} - \instantcost{t}{\optimalpolicy} \right].
\end{eqnarray}
{Moreover, let $\estpara{n}$ denote the LSE of the true dynamics parameter $\para_0$, using the sample $\left\{ x(i) \right\}_{i=0}^n, \left\{ u(j) \right\}_{j=0}^{n-1}$. In order to measure the precision of the learning procedure, we consider the error $\Mnorm{\estpara{n}-\para_0}{2}$. This study provides a non-asymptotic analysis of the performance of (causal and adaptive) control laws, addressing both aspects of learning and planning. The following two problems are rigorously addressed.
\begin{problem}[Regulation] \label{RegProb}
	For $\policy \in \causal$, determine high probability uniform regret bounds. That is, specify the sequence $\left\{\sequence{n}{\delta}\right\}_{n=1}^\infty$ and the integer $n_0\left(\delta\right)$ such that
	\begin{eqnarray*}
		\sup\limits_{n \geq n_0\left(\delta\right)} \sequence{n}{\delta} \regret{n}{\policy}<\infty,
	\end{eqnarray*}
	with probability at least $1-\delta$.
\end{problem}
\begin{problem}[Identification] \label{IdProb}
	Find the high probability learning accuracy that holds uniformly over time. That is, specify $\left\{\sequence{n}{\delta}\right\}_{n=1}^\infty,n_0\left(\delta\right)$, so that with probability at least $1-\delta$, the following holds:
	\begin{eqnarray*}
		\sup\limits_{n \geq n_0\left(\delta\right)} \sequence{n}{\delta}\Mnorm{\estpara{n}-\para_0}{}<\infty.
	\end{eqnarray*}
\end{problem}}

\section{General Input Perturbations} \label{general theory}
In this section, we study arbitrary causal control laws with perturbed input signal. The regret analysis which considers the effects of both the linear feedback gains, as well as the perturbation signals will be presented in Subsection~ \ref{exploitsubsection}. Subsequently, we address the identification performance in Subsection~\ref{exploresubsection}, and discuss how the structure of a causal control law determines the finite sample accuracy of learning $\para_0$. The main results, Theorem~\ref{RegretTheorem} and Theorem~\ref{EstTheorem}, both provide high probability bounds which hold uniformly over time.

\subsection{Regret Bound} \label{exploitsubsection}
We address Problem~\ref{RegProb} for the regret of causal policy $\policy=\left\{ L_t,v(t) \right\}_{t=0}^\infty$. In absence of a perturbation sequence (i.e., $v(t)=0$), it has been shown that $\regret{n}{\policy}$ scales as $\sum\limits_{t=0}^{n-1} \norm{ \left(\Lmatrix{\para_0} - L_t \right) x(t) }{2}^2$, as $n$ grows~\cite{faradonbeh2018optimality}. Theorem~\ref{RegretTheorem} generalizes the above result along the following directions. First, the input signals provided by the linear feedback gains $\left\{ L_t \right\}_{t=0}^\infty$ are perturbed with exogenous random signals $\left\{ v(t) \right\}_{t=0}^\infty$. Second, the upper bounds for the regret presented in this subsection hold uniformly over time. {Thus, the results in this subsection are still applicable, even if the time horizon of interacting with the system is not necessarily large. To proceed, we define the following energy quantities which will be used throughout the non-asymptotic analyses.}
\begin{deff} \label{signalbounds}
	For $\policy=\left\{ L_t,v(t) \right\}_{t=0}^\infty \in \causal$, suppose that $\left\{ \ubound{v_t} \right\}_{t=0}^\infty, \left\{ \ubound{\ell_t} \right\}_{t=0}^\infty$ are deterministic sequences satisfying $\norm{\left(\Lmatrix{\para_0}-L_t\right)x(t)}{2} \leq \ubound{\ell_t}$, $\norm{v(t)}{2} \leq \ubound{v_t}$, for all $t \geq 0$. Then, let $\totalbound{v}=\sum\limits_{t=0}^{n-1} \ubound{v_t}^2$, $\totalbound{\ell}=\sum\limits_{t=0}^{n-1} \ubound{\ell_t}^2$. Similarly, define $\ubound{w_t}, \ubound{x_t}$, as well as $\totalbound{w},\totalbound{x}$, for $w(t),x(t)$, respectively. Finally, denote $\ubound{w^*}=\max\limits_{1 \leq t \leq n} \ubound{w_t}$, and
	\begin{eqnarray*}
		\totalbound{\policy} = \left( \sum\limits_{t=0}^{n-1} \ubound{v_t}^2 \left( \ubound{\ell_t} + \ubound{x_t} \right)^2 \right)^{1/2}+ \left( \ubound{w^*}^2 \sum\limits_{t=0}^{n-1} \left( \ubound{\ell_t} + \ubound{v_t} \right)^2 \right)^{1/2}.
	\end{eqnarray*}
\end{deff}
The dependence of all terms $\totalbound{}$ on $n$ is suppressed for notational convenience. Intuitively, $\totalbound{y}$ corresponds to the deterministic upper bound for the energy of the signal $\left\{ y(t) \right\}_{t=0}^\infty$ up to time $n$. Subsequently, we establish a high probability regret bound in terms of the energy quantities defined above.
\begin{thrm}\label{RegretTheorem}
	Using Definition \ref{signalbounds} for $\policy= \left\{ L_t,v(t) \right\}_{t=0}^\infty$, there is a fixed constant $\ssconstant_1<\infty$ such that for all $\delta>0$,
	\begin{eqnarray*}
		\PP{\sup\limits_{n \geq 1} \frac{\regret{n}{\policy}}{\totalbound{v}+ \totalbound{\ell} + \totalbound{\policy} \log^{1/2} \left( n \delta^{-1} \right) } \leq \ssconstant_1} \geq 1 - \delta.
	\end{eqnarray*}
\end{thrm}
One can explicitly calculate $\ssconstant_1$ using the statements in the proof of Theorem \ref{RegretTheorem} in the \ifarxiv appendix\fi. To provide further intuition and outline the consequences of the above general result, we state the following corollary:
\begin{corol} \label{ExpRegretCor}
	Let $\E{\cdot}$ be the expectation with respect to the probability measures induced by $\policy$, $\left\{w(t)\right\}_{t=1}^\infty$. Then, with the notation of Theorem \ref{RegretTheorem}, we have 
	\begin{eqnarray*}
		\sup\limits_{n \geq 1} \frac{\E{\regret{n}{\policy}}}{\totalbound{v}+ \totalbound{\ell}} \leq \ssconstant_1.
	\end{eqnarray*}
\end{corol}

{Note that the expressions in the above finite-time regret bounds provide a novel decomposition as follows. First, the previous high probability upper bounds for the regret consist of summation of the terms $\left\{\ubound{\ell_t},\ubound{v_t}\right\}_{t=0}^{n-1}$~\cite{faradonbeh2017finite}, while in Theorem~\ref{RegretTheorem} and Corollary~\ref{ExpRegretCor}, all the terms $\totalbound{v},\totalbound{\ell},\totalbound{\policy}$ are in the form of sum-of-\emph{squares} of $\left\{\ubound{\ell_t},\ubound{v_t}\right\}_{t=0}^{n-1}$. For an adaptive regulator, $\ubound{\ell_t}$ shrinks as time goes by, since the decision-maker acquires more data to learn the dynamics. Hence, if the perturbation signal diminishes sufficiently fast, we get a significantly smaller regret bound compared to the existing results. For example, the magnitude of the lower bound for the regret changes from $n^{1/2}$~\cite{faradonbeh2017finite} to $\log n$~\cite{faradonbeh2018optimality}. Further, this regret specification sets the stage for Theorem~\ref{GreedyRates}, which establishes the finite time efficiency of a perturbed greedy regulator. In the sequel, we discuss other aspects of the result of Theorem~\ref{RegProb}.} 

Considering Theorem \ref{RegretTheorem} and Corollary \ref{ExpRegretCor}, the only different term is $\totalbound{\policy} \log^{1/2} \left( n \delta^{-1}\right)$. Thus, the stochastic behavior of $\policy$ is reflected by $\totalbound{\policy}$, and the other terms reflect the expected deviations from the optimal regulator $\optimalpolicy$. In other words, the cumulative deviation from the desired optimal trajectory consists of the systematic long lasting portion, as well as the spontaneous fluctuations. According to Corollary \ref{ExpRegretCor}, the former is reflected in the energy of the perturbation signal $\totalbound{v}$, and the energy of the instantaneous deviations in the feedback gains $\totalbound{\ell}$. The magnitude of the spontaneous fluctuations is with high probability determined by $\totalbound{\policy}$, which essentially reflects the interaction between $x(t),v(t), L_t-\Lmatrix{\para_0}$, as well as the interaction between $w(t+1), v(t), L_t-\Lmatrix{\para_0}$. So, $\totalbound{\policy}$ can be interpreted as the {\em fluctuation} energy. We will shortly compare the aforementioned involved quantities for different perturbation signals.

If there is no perturbation, $v(t)=0$, $\totalbound{\ell}$ is shown to be a tight asymptotic bound for $\regret{n}{\policy}$ \cite{faradonbeh2018optimality}. To compare with the result of Theorem \ref{RegretTheorem}, letting $\ubound{v_t}=0$ for all $t \geq 0$, suppose that the noise is uniformly bounded. Then, since $\sup\limits_{t \geq 1}\ubound{w_t}< \infty$, the fluctuation energy $\totalbound{\policy}$ can be replaced with $\totalbound{\ell}^{1/2}$. This leads to the non-asymptotic regret bound $\totalbound{\ell} + \totalbound{\ell}^{1/2} \log^{1/2} \left( n \delta^{-1} \right) $, which is a finite time counterpart for the asymptotic bound $\totalbound{\ell}$~\cite{faradonbeh2018optimality}. 

Next, we discuss a case of diminishing perturbations to compare the contributions of different factors toward the uniform bound of Theorem \ref{RegretTheorem}. Assuming $\sup\limits_{t \geq 1}\ubound{w_t}< \infty$, let $\left\{v(t)\right\}_{t=0}^\infty$ be a diminishing signal such that $\sup\limits_{t \geq 0} t^{\alpha_1}\ubound{v_t}<\infty$, for some $\alpha_1>0$. Further, let the deviation from the optimal feedback satisfy
\begin{eqnarray*}
	\sup\limits_{t \geq 0} t^{\alpha_2}\Mnorm{\Lmatrix{\para_0}-L_t}{2} < \infty,
\end{eqnarray*} 
for some $\alpha_2>0$. Since the above policy stabilizes the system~\cite{faradonbeh2018stabilization}, the state signal is uniformly bounded as well. Then, Theorem \ref{RegretTheorem} provides the following uniform regret bound
\begin{eqnarray*}
	{n^{1- 2\alpha_1 \wedge 2\alpha_2} + n^{1/2-\alpha_1 \wedge \alpha_2} \log^{1/2} \left( n \delta^{-1} \right)}.
\end{eqnarray*}
Similarly, it is straightforward to show that uniform boundedness of the noise vectors and the state signal leads to the bound
\begin{eqnarray*}
	{ \totalbound{v}+ \totalbound{\ell} + \left( \totalbound{v}+ \totalbound{\ell} \right)^{1/2} \log^{1/2} \left( n \delta^{-1} \right)}.
\end{eqnarray*}
Therefore, the perturbation signal does not leads to an increase in the high probability regret bound as long as it diminishes as fast as ${\Lmatrix{\para_0}-L_t}$. For the special case of $\alpha_1 \wedge \alpha_2=1/2$, the regret is of the order
\begin{eqnarray*}
	{\log n + \log^{1/2} n \log^{1/2}\delta^{-1}},
\end{eqnarray*}
which is also the information theoretic lower bound~\cite{faradonbeh2018optimality}.

{Finally, Theorem \ref{RegretTheorem} shows that the non-asymptotic high probability uniform bound of the regret is memoryless~\cite{faradonbeh2018optimality}. That being said, if hypothetically the operator stops using perturbed inputs or deviated feedback gains, the regret bound of Theorem~\ref{RegretTheorem} freezes and does not grow anymore. Therefore, Theorem~\ref{RegretTheorem} and Corollary~\ref{EstTheorem} indicate that the adaptive decision-maker can revoke the sub-optimal control inputs in the past time. In other words, the history of the non-optimal actions previously taken, does not significantly influence the future trajectory of the dynamical system (memorylessness).}

\subsection{Identification Bound} \label{exploresubsection}
Next, we address Problem~\ref{IdProb} by analyzing the effect of the input perturbation for learning the unknown dynamics parameter $\para_0$. Based on the linear data generation mechanism in~\eqref{systemeq1}, a natural estimation procedure for $\para_0$ is least squares. That is, to regress every state sample $x(t+1)$ on the previous samples of input $u(t)$ and state $x(t)$, to get the following LSE:
\begin{eqnarray} \label{LSE}
	\estpara{n} &=& \arg\min\limits_{\para \in \R^{p \times q}} \sum\limits_{t=0}^{n-1} \norm{x(t+1)- \para \begin{bmatrix}
			x(t) \\ u(t)
		\end{bmatrix}}{2}^2.
\end{eqnarray}
In the statistical analyses of the accuracy of $\estpara{n}$, the associated stochastic process $\left\{ w(t) \right\}_{t=1}^\infty$, as well as the exogenous one $\left\{ v(t) \right\}_{t=0}^\infty$, play an important role. So, we use the following definition which uses the second moments of the above stochastic processes. 
\begin{deff} \label{mineigdefn}
	Suppose that $\policy=\left\{ L_t,v(t) \right\}_{t=0}^\infty \in \causal$. Then, let $\covmat{w} = \sum\limits_{t=1}^{n-1} \E{w(t)w(t)'}$, $\covmat{v} = \sum\limits_{t=0}^{n-2} \E{v(t)v(t)'}$, and $\lbound{\lambda_n} = \eigmin{\covmat{w}} \wedge \eigmin{\covmat{v}}$. Further, letting $D_0=A_0+B_0 \Lmatrix{\para_0}$, and
	\begin{eqnarray*}
		\ubound{y_t}=\Mnorm{B_0}{2}\ubound{v_t} +\ubound{w_{t+1}} + 2\Mnorm{B_0}{2} \ubound{\ell_t} + 2\Mnorm{D_0}{2} \ubound{x_t},
	\end{eqnarray*}
	define the following quantities similar to Definition \ref{signalbounds}:\\
	\begin{eqnarray*}
		\totalbound{1} &=& \left(\sum\limits_{t=0}^{n-2} \ubound{y_t}^2 \left( \Mnorm{B_0}{2}\ubound{v_t} +\ubound{w_{t+1}}\right)^2\right) ^{1/2}, \\
		\totalbound{2} &=& \left(\sum\limits_{t=0}^{n-2}  \ubound{v_t}^2 \left( 2 \left[ 1+\Mnorm{\Lmatrix{\para_0}}{2} \right] \ubound{x_t} + 2\ubound{\ell_t} + \ubound{v_t} \right)^2\right)^{1/2}.
	\end{eqnarray*}
\end{deff} 
Note that $\covmat{w},\covmat{v},\totalbound{1},\totalbound{2}$ all depend on $n$. In order to interpret $\totalbound{1},\totalbound{2}$, apply the causal policy $\policy=\left\{ L_t,v(t) \right\}_{t=0}^\infty$ to the dynamical model in \eqref{systemeq1}. Then, the closed-loop dynamics becomes $x(t+1)=D_tx(t)+B_0v(t)+w(t+1)$, where $D_t=A_0+B_0L_t$. Thus, the noise in the evolution of the system is $B_0v(t)+w(t+1)$, which indicates that $\totalbound{1}$ reflects a deterministic upper bound for the interaction energy between the observed signals $x(t)$, and the unobserved noise $B_0v(t)+w(t+1)$. Similarly, $\totalbound{2}$ bounds the interaction energy between the signal $x(t)$ and the perturbation $v(t)$. 

In order to accurately identify the true parameter $\para_0$, $\lbound{\lambda_n}$ cannot be very small. Broadly speaking, the magnitude of $\lbound{\lambda_n}$ determines the extend to which both the perturbation and the noise processes {\em excite} all coordinates of the matrix $\para_0$. So, the precision of the learning procedure in \eqref{LSE} highly depends on $\lbound{\lambda_n}$. To establish a uniform upper bound for learning accuracy, we assume the followings for the quantities in Definition \ref{mineigdefn}:
\begin{eqnarray} 
	\eigmin{\covmat{w}} &\geq& 4 \totalbound{1} \log^{1/2} \left( 8p \delta^{-1} \right), \label{minSSforw} \\
	\eigmin{\covmat{v}} &\geq& 4 \totalbound{2} \log^{1/2} \left( 8q \delta^{-1} \right). \label{minSSforv}
\end{eqnarray}
{Intuitively, the above inequalities generalize the persistent excitation condition~\cite{green1985persistence} to cases with diminishing (i.e., non-persistent) perturbations. Later on, we will show that one can easily satisfy \eqref{minSSforw}, \eqref{minSSforv}. Further, writing down the closed-loop dynamics, $\eigmin{\covmat{w}}$ denotes the minimum energy of the noise sequence $B_0 v(t)+w(t)$. Going back to the interpretation of $\totalbound{1}$ after definition \ref{mineigdefn}, the condition \eqref{minSSforw} indeed states that the noise energy is larger than the interaction energy (of the noise and the signal $x(t)$), with high probability. Therefore, the interaction is not powerful enough to prevent the noise from exciting all coordinates of the unknown transition matrix $A_0$. Similar explanation for \eqref{minSSforv} states that the excitation of $B_0$ by the perturbation signal is not masked by its interaction with the state signal.} These are formally presented in the following result. 
\begin{thrm} \label{EstTheorem}
	Using the notation of Definitions \ref{signalbounds}, \ref{mineigdefn}, with probability at least $1-\delta$ we have
	\begin{eqnarray*}
		\sup \frac{ \lbound{\lambda_n} \Mnorm{\estpara{n}-\para_0}{2}^2}{ \ubound{w^*}^2 \log \left[\left(\totalbound{\ell} + \totalbound{x} + \totalbound{v}\right) \delta^{-1}\right]} \leq \ssconstant_2 ,
	\end{eqnarray*}	
	where the supremum is over all $n \geq 1$ satisfying \eqref{minSSforw}, \eqref{minSSforv}.
\end{thrm}
The constant $\ssconstant_2$ depends on the true dynamics parameter $\para_0$, and can be extracted from the proof in the \ifarxiv appendix\fi. Next, to discuss Theorem \ref{EstTheorem}, suppose that the covariance matrices of the noise vectors are bounded from below: $\inf\limits_{n \geq 1} n^{-1} \eigmin{\covmat{w}} >0$. Further, for the perturbation process assume that
\begin{eqnarray*}
	\inf\limits_{t \geq 1} t^{\alpha_1} \eigmin{\E{v(t)v(t)'}} > 0, \:\:\:\:\:\: \sup\limits_{t \geq 1} t^{\alpha_2} \ubound{v_t} < \infty,
\end{eqnarray*}
for some $0 \leq \alpha_1 <1 $, and $0 \leq \alpha_2 \leq 1$. Hence, we obtain $\inf\limits_{n \geq 1} n^{\alpha_1-1} \lbound{\lambda_n} >0$. Further, similar to the discussion after Theorem \ref{RegretTheorem}, suppose that both the noise and the state signal are uniformly bounded (e.g., $\ubound{\ell_t}$ is diminishing). Assuming $\alpha_1-\alpha_2 < 1/2$, since
\begin{eqnarray*}
	\sup\limits_{n \geq 1} n^{-1/2} \totalbound{1} < \infty,
	\:\:\:\:\:\: \sup\limits_{n \geq 1} n^{\alpha_2 - 1/2} \totalbound{2} < \infty,
\end{eqnarray*}
the inequalities in \eqref{minSSforw}, \eqref{minSSforv} are satisfied as long as $n$ is up to a constant factor at least $\left( - \log \delta \right)^{1/\left(1-2 \alpha_1+ 2\alpha_2 \right)}$. Then, Theorem \ref{EstTheorem} implies the identification bound $n^{\alpha_1/2 - 1/2} \log^{1/2} \left(n \delta^{-1}\right)$ for uniform learning accuracy of estimating the unknown parameter $\para_0$. Hence, the identification accuracy is basically determined by the diminishing rate of the perturbation signal. Moreover, for the case of persistent perturbation $\alpha_1=\alpha_2=0$, LSE in \eqref{LSE} achieves the optimal uniform learning rate of $n^{-1/2} \log^{1/2} \left(n \delta^{-1}\right)$. Note that in comparison to the analysis presented in the previous subsection, a larger perturbation leads to more accurate learning together with larger deviations from the optimal cost, and vice versa. Balancing this trade-off is the main challenge for adaptive policies, and will be addressed in the sequel. Finally, for $\alpha_1=1/2$, the result extends the existing asymptotic identification rates~\cite{faradonbeh2018optimality}. 

\section{Perturbed Greedy Regulator} \label{CE section}
Next, we analyze the non-asymptotic regret of perturbed greedy policies, as well as their identification errors. For this purpose, we state a fairly general condition about the probabilistic properties of the noise process, and present an adaptive regulator in Algorithm~\ref{algo}. The subsequent contents consist of establishing that Algorithm~\ref{algo} addresses the main dilemma of reinforcement learning; balancing the exploration and the exploitation. In this section, we assume the following about the noise process $\left\{ w(t) \right\}_{t=1}^\infty$ in~\eqref{systemeq1}.
\begin{assump} \label{tailcondition}
	The noise has a sub-Weibull distribution. That is, for some fixed $\tailconst, \tailcoeff , \tailexp >0$, and for all $t \geq 1$, and $\eta >0$, we have 
	\begin{eqnarray*}
		\PP{ \norm{w(t)}{2} > \eta} \leq \tailconst \: \exp \left(-{\tailcoeff}^{-1}{\eta^{\tailexp}}\right)
	\end{eqnarray*}
	Further, we assume that the covariance matrices of the noise vectors satisfy
	\begin{eqnarray*}
		\inf\limits_{n \geq 1} n^{-1} \sum\limits_{t=1}^{n} \eigmin{\E{w(t)w(t)'}} \geq \sigma_0 >0.
	\end{eqnarray*}
\end{assump}
Note that positive definiteness of the covariance matrices $\E{w(t)w(t)'}$ is sufficient for the second part of Assumption \ref{tailcondition}, but is not necessary. Regarding the first part of Assumption \ref{tailcondition}, sub-Weibull distributions are remarkably general in learning theory when analyzing the finite-time performance~\cite{faradonbeh2018finite}. In fact, they encompass a wide range of distributions being commonly used in the literature, such as uniformly bounded ($\tailexp = \infty$), sub-Gaussian ($\tailexp=2$), and sub-Exponential ($\tailexp=1$) random vectors, as well as heavy-tail distributions for which moment generating functions do not exist ($\tailexp<1$). Later on, we will see that the exponent $\tailexp$ which determines the decay rate in the probability distributions of the noise vectors, plays a crucial role in the non-asymptotic analyses for both the regret and the identification error. Note that the noise distributions are not required to have densities or continuous cumulative distribution functions.

\begin{algorithm}
	\caption{{: Perturbed Greedy Regulator} } \label{algo}
	\begin{algorithmic}
		\State Inputs: $\rrate >1$, $0 < r \lbound{\ssconstant}< \ubound{\ssconstant} < \infty$
		\State Let $\Lmatrix{\estpara{0}} \in \R^{r \times p}$ be a stabilizer 
		\For{$m=0,1,2,\cdots$}
		\While{$t < \rrate^m$}
		\State Draw perturbation $v(t)$ according to \eqref{PerturbDesign}
		\State Apply input $u(t)=\Lmatrix{\estpara{t}} x(t)+ v(t)$ 
		\State $\estpara{t+1}=\estpara{t}$
		\EndWhile
		\State Update the estimate $\estpara{t}$ by \eqref{LSE}
		\EndFor
	\end{algorithmic}
\end{algorithm}
The pseudo-code of the perturbed greedy regulator is provided in Algorithm~\ref{algo}. It is an episodic algorithm; learning of the unknown parameters is deferred until sufficiently many observations have been collected. More precisely, Algorithm~\ref{algo} updates the parameter estimates only at the end of epochs of exponentially growing length. This lets the solution of Riccati equation~\eqref{ricatti2} be efficiently used by preventing unnecessary computations. Note that numerical computation of the optimal feedback gain in~\eqref{ricatti1} is practically nontrivial~\cite{fazel2018global}. 

\begin{figure} 
	\centering
	\scalebox{0.40}
	{\includegraphics{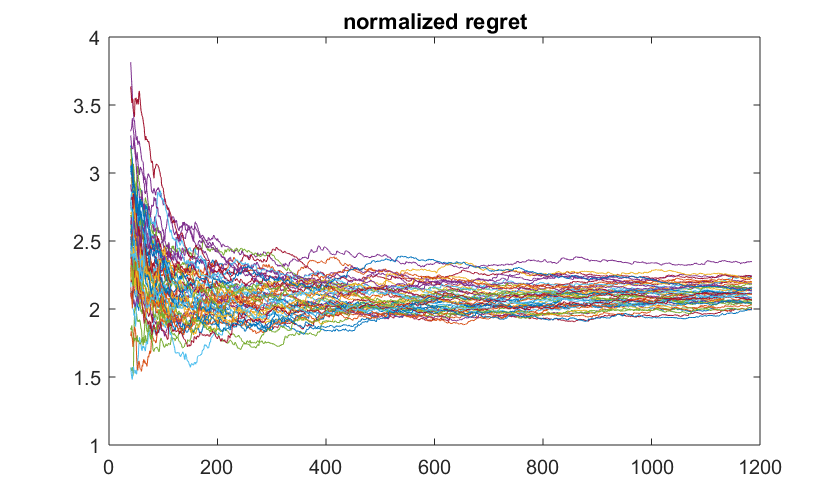}} 
	\scalebox{.40}
	{\includegraphics{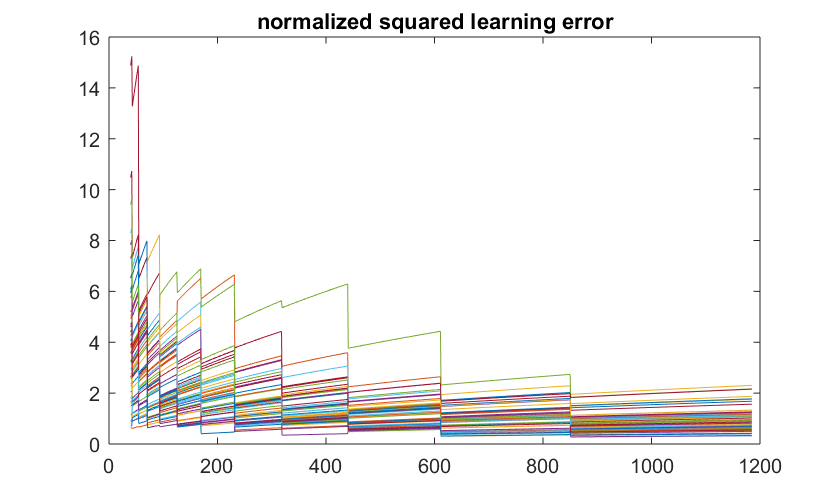}} 
	\caption{{Performance of Algorithm~\ref{algo}: normalized regret $\left({n^{-1/2} \log^{-2} n} \right){\regret{n}{\adaptivepolicy}}$ vs $n$ (top), and normalized squared learning error $n^{1/2} {\Mnorm{\estpara{n}-\para_0}{}^2}$ vs $n$ (bottom).}}
	\label{Fig1}
\end{figure}
Formally, the algorithm starts with an arbitrary initial stabilizing feedback matrix $\Lmatrix{\estpara{0}}$. An algorithmic procedure for computing $\Lmatrix{\estpara{0}}$ in finite time is available in the literature, together with theoretical performance guarantees for stabilization sets~\cite{faradonbeh2018stabilization}. Therefore, we can assume that the closed-loop system remains stable during the execution of Algorithm~\ref{algo}. At each time $t=0, 1, 2, \cdots$, Algorithm~\ref{algo} applies the adaptive control input $u(t)=\Lmatrix{\estpara{t}} x(t)+ v(t)$. The design of the perturbation signals $v(t)$ will be discussed shortly. Further, the lengths of the epochs are determined by the rate $\rrate>1$ as follows. If $t$ satisfies $t= \lfloor \rrate^m \rfloor$ for some $m=0,1,\cdots$, Algorithm~\ref{algo} tunes the regulator by finding $\estpara{t}$ according to \eqref{LSE}. Otherwise, for $t \neq \lfloor \rrate^m \rfloor$, the learning step will be skipped and no update occurs: $\estpara{t}=\estpara{t-1}$. {The perturbation signals during each epoch $m \geq 1$ are wide-sense stationary (homoscedastic) bounded random signals with positive definite covariance matrices; for all $\rrate^m \leq t < \rrate^{m+1}$, we have $\E{v(t)v(t)'}=\covmat{m}$, and $\norm{v(t)}{2}< \ubound{v_m}$, such that for some {arbitrary} universal constants $\lbound{\ssconstant}>0$, $\ubound{\ssconstant}<\infty$,
\begin{eqnarray} \label{PerturbDesign}
	\lbound{\ssconstant} < m^{-2}\rrate^{m/2} \eigmin{\covmat{m}} \leq m^{-2}\rrate^{m/2} \ubound{v_m}^2 < \ubound{\ssconstant}.
\end{eqnarray}
In general, any sequence of mean-zero independent random vectors satisfying~\eqref{PerturbDesign} can be used as the perturbation signal. Note that we need $r \lbound{\ssconstant} \leq \ubound{\ssconstant}$, since
\\$r \eigmin{\covmat{m}} \leq \tr{\E{v(t)v(t)'}} = \E{v(t)'v(t)} \leq \ubound{v_m}^2$.} 

Although larger values of $\rrate$ make the epochs longer, and thus fewer updates of the parameter estimates occur, the rates of regulation and identification do not depend on the magnitude of $\rrate$. Intuitively, it is because Algorithm~\ref{algo} utilizes all the random perturbation signals during each epoch when learning the dynamics parameter at the end of that epoch. The following theorem addresses the uniform finite time rates for the above regulator.
\begin{thrm} \label{GreedyRates}
	Suppose that $\adaptivepolicy$ is the adaptive regulator of Algorithm~\ref{algo}. Let $\estpara{n}$ be the parameter estimate at time $n$. Then, with probability at least $1-\delta$ we have
	\begin{eqnarray*}
		\sup\limits_{n \geq n_0} \frac{\regret{n}{\adaptivepolicy}}{n^{1/2} \log^{2 \vee 4/\tailexp} \left(n \delta^{-1} \right)} &\leq& \ssconstant_3, \\
		\sup\limits_{n \geq n_0} \frac{ \Mnorm{\estpara{n}-\para_0}{2}^2}{n^{-1/2} \log^{2/\tailexp-1} \left(n \delta^{-1} \right) } &\leq& \ssconstant_4,
	\end{eqnarray*}
	where $n_0 \geq \ssconstant_0 \left(\log^{2+4/\tailexp} \delta^{-1}\right) \left(\log\log \delta^{-1}\right)$. 
\end{thrm}
The fixed constants $\ssconstant_0,\ssconstant_3,\ssconstant_4$ depend on the truth $\para_0$, the noise parameters $\tailcoeff,\tailconst,\sigma_0$, and the constants $\rrate,\lbound{\ssconstant},\ubound{\ssconstant}$ in~\eqref{PerturbDesign}. {Using the notation of Theorem~\ref{EstTheorem}, $\lbound{\lambda_n}$ scales linearly with $\lbound{\ssconstant}$. Thus, the constant $\ssconstant_4$ in Theorem~\ref{GreedyRates} is proportional to $\lbound{\ssconstant}^{-1}$, as well as the quantity $\totalbound{\ell}$ in Theorem~\ref{RegretTheorem} (see~\eqref{LipschitzL}). Further, with the notation of Theorem~\ref{RegretTheorem}, $\totalbound{v}$ scales linearly with $\ubound{\ssconstant}$. Therefore, the constant $\ssconstant_3$ in Theorem~\ref{GreedyRates} scales proportional to $\ubound{\ssconstant}+\lbound{\ssconstant}^{-1}$.} Figure~\ref{Fig1} depicts the performance of Algorithm~\ref{algo} for the dynamics and cost matrices in~\eqref{simulationmatrices}, {and $\rrate=1.2,\lbound{\ssconstant}=1,\ubound{\ssconstant}=10$. The distribution of the random vectors $v(t)$ is Gaussian, truncated to satisfy~\eqref{PerturbDesign}. Finally, the initial stabilizer $\Lmatrix{\estpara{0}}$ is provided by applying the stabilization algorithm of Faradonbeh et al.~\cite{faradonbeh2018stabilization} to the first $17$ samples; $\left\{ x(t) \right\}_{t=0}^{16}, \left\{ u(t) \right\}_{t=0}^{15}$.} The curves in Figure~\ref{Fig1} correspond to different replicates, and fully reflect the statement of Theorem~\ref{GreedyRates}.
{\begin{table*}
	{\small \begin{eqnarray} \label{simulationmatrices}
		A_0=\begin{bmatrix}
			0.13&    0.35&   -0.26\\
			-1.34&   -0.30&   -1.75\\
			1.18&         0&   -0.29
		\end{bmatrix}, \:\:\:
	B_0=\begin{bmatrix}
		-0.83&   -0.53&    0.52\\
		-0.98&   -2.00&         0\\
		-1.16&    0.96&   -0.04
	\end{bmatrix}, \:\:\:
	Q=\begin{bmatrix}
		0.79&   -0.15&    0.09\\
		-0.15&    0.60&   -0.04\\
		0.09&   -0.04&    0.61
	\end{bmatrix}, \:\:\:
	R=\begin{bmatrix}
		0.52&   -0.06&   -0.07\\
		-0.06&    0.39&   -0.04\\
		-0.07&   -0.04&    0.67
	\end{bmatrix}. \:\:\:\:\:\:\:
	\end{eqnarray}}
\end{table*}}

{Subsequently, we compare Theorem~\ref{GreedyRates} with the relevant asymptotic results obtained without input perturbation. The comparisons focus on the performance of regulation and identification of Algorithm~\ref{algo} vis-a-vis  those of previously proposed adaptive regulators known as Randomized Certainty Equivalence (RCE) and Thompson Sampling (TS)~\cite{faradonbeh2018optimality}. The former policy consists of computing the LSE of the true parameter at the end of each epoch, and then adding a random matrix to it. The latter policy adopts a Bayesian approach, and approximates $\para_0$ with draws from Gaussian posteriors at the end of every epoch. Although the presented asymptotic bounds for regret and learning errors of RCE and TS are similar to those of Algorithm~\ref{algo} (see~\cite{faradonbeh2018optimality}), their non-asymptotic performance deteriorates due to fluctuations in the state trajectory. Intuitively speaking, when the time period of interacting with the system is finite, polynomial functions of $n\delta^{-1}$ appear in the expressions of both regret and learning error. Therefore, the potential stochastic fluctuations are neither sufficiently {\em scarce} (because of $\delta^{-1}$), nor {\em small} enough (because of $n$). Note that the results of Theorem~\ref{GreedyRates} are stronger than the existing analyses also because of providing uniform bounds (the previous results address `$\limsup$' instead of `$\sup$').} 

\begin{figure}[t!] 
	\centering
	\scalebox{.40}
	{\includegraphics{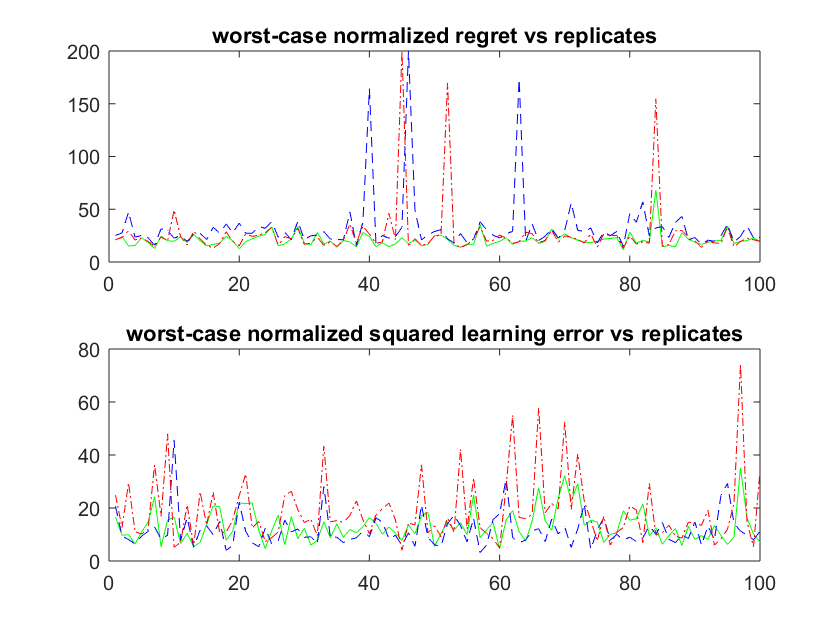}} 
	\caption{{Performance of Algorithm~\ref{algo} (\textcolor{green}{green}) compared to RCE (\textcolor{blue}{blue}), and TS (\textcolor{red}{red})~\cite{faradonbeh2018optimality} for 100 replicates: worst normalized regret $\sup\limits_{n} n^{-1/2} \regret{n}{\adaptivepolicy}$ (top), and worst normalized squared learning error $\sup\limits_{n} n^{1/2} {\Mnorm{\estpara{n}-\para_0}{}^2}$ (bottom).}}
	\label{worstvsreplicate}
\end{figure}
\begin{figure}[t!] 
	\centering
	\scalebox{.40}
	{\includegraphics{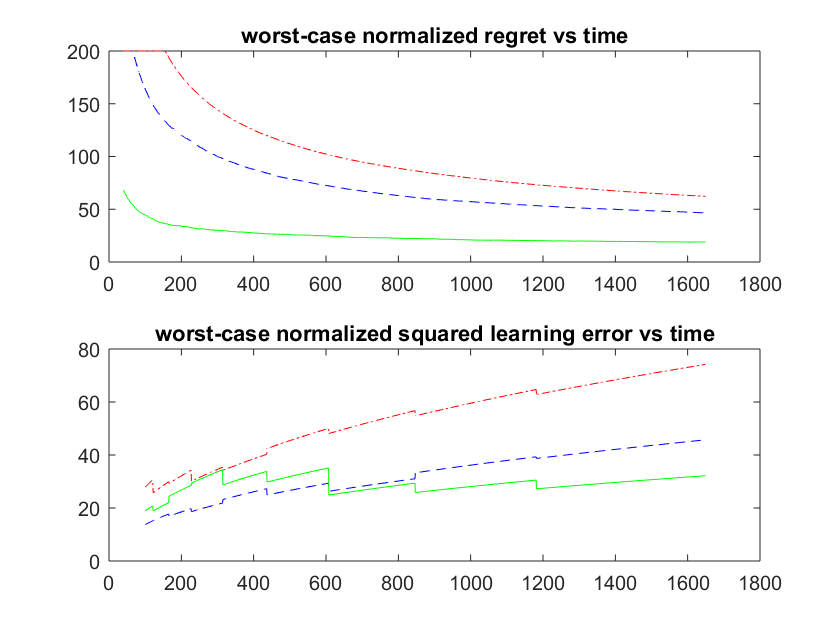}} 
	\caption{{Upper bounds of regulation and identification obtained by 100 replicates of Algorithm~\ref{algo} (\textcolor{green}{green}), RCE (\textcolor{blue}{blue}), and TS (\textcolor{red}{red})~\cite{faradonbeh2018optimality}: the largest values of the normalized regret $n^{-1/2} {\regret{n}{\adaptivepolicy}}$ (top), and the normalized squared learning error $n^{1/2} {\Mnorm{\estpara{n}-\para_0}{}^2}$ (bottom), both as functions of $n$.}}
	\label{worstvstime}
\end{figure}
{These improvements are visualized for the system of the previous numerical example.} In Figure~\ref{worstvsreplicate}, we report the uniform upper bound of the normalized regret and the normalized learning error for 100 replicates. Clearly, the fluctuations of Algorithm~\ref{algo} are significantly smaller that the counterparts of RCE and TS. In Figure~\ref{worstvstime}, the normalized performance quantities are calculated versus time, and the largest among the 100 replicates is being plotted. The non-asymptotic improvement of the perturbed greedy regulator over RCE and TS can be seen in both figures. A recent study provides an analogous regret bound for single dimensional systems, wherein at every time step the parameter estimate is being updated and perturbed by diminishing Gaussian random matrices~\cite{abeille2018improved}. The computational cost of the latter regulator is exponentially larger than that of Algorithm \ref{algo}, since Riccati equations need to be repeatedly solved~\cite{abeille2018improved}.

\subsection{Partial Uncertainty of Parameters} \label{CE with side}
Next, we study the effect of restricted uncertainty about the true dynamical parameter $\para_0$ on the performance of adaptive regulators. In fact, the context of the plant under consideration is capable of providing side information about $\para_0$. Examples of such information include the structural connectivity of networks which determines the support of $\para_0$, and artificially designed systems where the construction imposes restriction on possible sets of $p \times q$ dynamics matrices. Further, in the systems with longer memory where the current state depends on multiple previous time steps, the transition matrix $A_0$ has a particular structure once the dynamics is written in the form of~\eqref{systemeq1}~\cite{guo1991aastrom,faradonbeh2018finite}. Therefore, the adaptive operator can restrict the learning of the unknown dynamics parameter to an specific subset of $\R^{p \times q}$.

In the sequel, we show that such additional information can lead to significant improvements in the regret bounds. Although such situations are investigated for the settings with finitely many possible control inputs, non-asymptotic analyses are not currently available for LQ systems. In fact, for general Markov Decision Processes (MDP) it is known that the regret is modulo a constant factor at least $n^{1/2}$~\cite{jin2018q}. However, if one knows that the underlying MDP has a \emph{distinguishability} property, the regret can be of order $\log n$~\cite{tewari2008optimistic}. A similar situation holds for Multi-Armed Bandits (MAB)~\cite{bastani2017mostly}. 

To proceed, we define the following identifiability condition as the analogue of the distinguishability condition in MDPs. Intuitively, in the identifiable systems, the optimal feedback gain $\Lmatrix{\para_0}$ can be learned \emph{as accurately as} the closed-loop transition matrix $A_0+B_0 \Lmatrix{\estpara{t}}$. 
\begin{deff} \cite{faradonbeh2018optimality} \label{Sidentifiability}
	Assuming $\para_0 \in \tempparaspace{0}$ for some $\tempparaspace{0} \subset \R^{p \times q}$, $\para_0$ is called identifiable if 
	\begin{eqnarray} \label{Sideq}
	\sup\limits_{\para_1,\para_2 \in \tempparaspace{0}} \frac{\Mnorm{\Lmatrix{\para_2} - \Lmatrix{\para_0}}{}}{\Mnorm{\left(\para_2 - \para_0\right) \extendedLmatrix{\para_1}}{}} < \infty,
	\end{eqnarray}
	where $\extendedLmatrix{\para}= \left[I_p , \Lmatrix{\para}'\right]'$.
\end{deff}
\begin{remark}
	It has been established that the feedback difference $\Lmatrix{\para}-\Lmatrix{\para_0}$ is dominated by the parameter difference $\para-\para_0$, modulo a constant factor~\cite{faradonbeh2017finite}:
	\begin{eqnarray} \label{LipschitzL} 
	\sup\limits_{\para: \Mnorm{\Kmatrix{\para}}{2} \leq \ssconstant_K } \frac{\Mnorm{\Lmatrix{\para} - \Lmatrix{\para_0}}{2}}{\Mnorm{\para - \para_0}{2}} \leq \ssconstant_L,
	\end{eqnarray}
	where $\ssconstant_L < \infty$ is a fixed constant determined by $\para_0$ and $\ssconstant_K<\infty$. Note that \eqref{Sideq} is stronger than \eqref{LipschitzL} since in general, the identifiability condition does not need to hold. Indeed, one can have $\Lmatrix{\para_2} \neq \Lmatrix{\para_0}$, but at the same time $\para_2 \extendedLmatrix{\para_1} = \para_0 \extendedLmatrix{\para_1}$~\cite{faradonbeh2018optimality}. 
\end{remark} 
{An extensive discussion of concrete examples of $\tempparaspace{0}$ is provided in the work of Faradonbeh et al. \cite{faradonbeh2018optimality}. In general, $\tempparaspace{0}$ can be a manifold (or a finite union of manifolds) induced by different conditions such as the rank or the sparsity of the dynamics matrices. In the sequel, we briefly discuss four examples. The first condition is the support of the dynamics matrices. Indeed, letting $\mathcal{I} \subset \left\{ (i,j): 1 \leq i \leq p, 1 \leq j \leq q \right\}$, assume that $\para$ belongs to $\tempparaspace{0}$, if and only if the coordinate $(i,j)$ of $\para$ is zero for all $(i,j) \notin \mathcal{I}$. If the cardinality of $\mathcal{I}$ is not larger than $ p^2$, $\tempparaspace{0}$ can satisfy \eqref{Sideq} \cite{faradonbeh2018optimality}. The second condition is based on sparsity of the true parameter. That is, if $\para_0$ is known to have at most $p^2$ non-zero entries, $\tempparaspace{0}$ is contained in the finite union of the subspaces satisfying the aforementioned support condition. Third, the rank of $\para_0$ can be used in the design of the adaptive policies as follows. The set of $p \times q$ matrices of rank at most $d$ is a finite union of manifolds of dimension at most $d \left( p+q-d \right)$~\cite{shalit2012online}, which can satisfy \eqref{Sideq} if $d \left( p+q-d \right) \leq p^2$~\cite{faradonbeh2018optimality}. Finally, one can utilize the knowledge about a linear subspace that $\para_0$ belongs to. Suppose that $\para_0$ is known to belong to a subspace of $\R^{p \times q}$. If the above subspace is of dimension at most $p^2$, it can satisfy the identifiability condition~\eqref{Sideq}~\cite{faradonbeh2018optimality}.}

{The quantity $p^2$ appearing in the above examples is actually determined by the learning invariant manifold $\nullspace{}$. Indeed, $\nullspace{}$ contains the dynamics matrices $\para$ for which the closed-loop matrix is indistinguishable from the truth $\para_0$; that is, $\para \extendedLmatrix{\para_0}=\para_0 \extendedLmatrix{\para_0}$. It has been established that if $\tempparaspace{0} \cap \nullspace{}=\left\{ \para_0 \right\}$, then $\tempparaspace{0}$ satisfies~\eqref{Sideq}~\cite{faradonbeh2018optimality}. Since $\nullspace{}$ is a linear subspace of dimension $pr$, we restricted the dimension of $\tempparaspace{0}$ to $pq -pr=p^2$. More technical details can be found in existing studies on learning invariant manifolds, optimality level-set manifolds, and their effect on the performance of adaptive LQRs~\cite{polderman1986necessity,polderman1986note,faradonbeh2018optimality}.}

In order to modify Algorithm~\ref{algo} for the systems with the side information $\tempparaspace{0}$, we replace \eqref{PerturbDesign} with $\rrate^{m} \ubound{v_m}^2 < \ubound{\ssconstant}$; i.e., the perturbation signals are smaller (and diminish faster) than \eqref{PerturbDesign}, and the covariance matrix does not need to be positive definite. Further, the uncertain dynamics parameter $\para_0$ will be learned through the following least squares procedure on the identifiability set $\tempparaspace{0}$:
\begin{eqnarray} \label{IDLSE}
	\estpara{n} &=& \arg\min\limits_{\para \in \tempparaspace{0}} \sum\limits_{t=0}^{n-1} \norm{x(t+1)- \para \begin{bmatrix}
		x(t) \\ u(t)
		\end{bmatrix}}{2}^2.
\end{eqnarray}
All other parts of the adaptive regulator, such as the lengths of the epochs, remain the same as in Algorithm~\ref{algo}. We conclude this section with the next result that provides a smaller regret bound vis-a-vis that of Theorem~\ref{GreedyRates}.
\begin{thrm} \label{IdGreedyRates}
	Let $\para_0$ be identifiable, and $\adaptivepolicy$ be the above adaptive regulator. Then, with probability at least $1-\delta$,
	\begin{eqnarray*}
		\sup\limits_{n \geq n_0} \frac{\regret{n}{\adaptivepolicy}}{\log n \log^{1+2/ \tailexp} \left(n \delta^{-1} \right)} &\leq& \ssconstant_5,
	\end{eqnarray*}
	where $n_0$ is defined in Theorem \ref{GreedyRates}, and the fixed constant $\ssconstant_5 <\infty$ depends on $\para_0$.
\end{thrm}

\section{Concluding Remarks}
We analyzed the performance of adaptive control policies with perturbed input for multidimensional LQ systems. Indeed, we established non-asymptotic results for the high probability regret bound (Theorem \ref{RegretTheorem}), as well as the learning accuracy (Theorem \ref{EstTheorem}), which hold uniformly over time. Leveraging the developed general theory, we showed that a greedy adaptive regulator with suitably designed input perturbation provides (nearly) square root regret with respect to time (Theorem \ref{GreedyRates}), and outperforms previously available reinforcement learning algorithms in a number of ways. 

Specifically, the presented perturbed greedy regulator does not need to know the spectral properties of the closed-loop transition matrix, and the noise statistics. Further, the computationally intractable steps being used by optimism-based adaptive policies are not required. Finally, input perturbation efficiently regulates the trajectory of the system by preventing fluctuations due to the stochastic nature of the dynamics. We also discussed the situations where the regret is of a logarithmic magnitude, assuming that the adaptive operator has access to some information (such as the support or the rank) of the unknown dynamics matrices (Theorem~\ref{IdGreedyRates}).

Extending the presented framework to the case of imperfect observations where a linear transformation of the state signal is observed with some noise is an interesting direction for future work. Moreover, addressing the problem of adaptive regulation for large scale systems (e.g., networks) in a possibly high dimensional setting such as sparse or low rank dynamical models is of interest for further investigation. Finally, settings involving time varying dynamics (e.g., switching systems and Markov Jump Processes) constitute topics for future studies.

\ifarxiv
\appendix
\newpage
\section{Proofs}
\subsection{\bf Proofs of Theorem \ref{RegretTheorem} and Corollary \ref{ExpRegretCor}}
	We start by extending the policy decomposition technique introduced in \cite{faradonbeh2018optimality}. For a policy $\policy=\left\{ L_t, v(t)\right\}_{t=0}^{n-1}$ and a fixed $n \geq 1$, define the sequence of policies $\policy_0, \cdots, \policy_n$ according to $\policy$:
	\begin{eqnarray*}
	\policy_i: \begin{cases}
		u(t)=L_t x(t)+v(t) & t<i\\
		u(t)=\Lmatrix{\para_0}x(t) & t \geq i.
	\end{cases}
	\end{eqnarray*} 
	Indeed, for $0 \leq i \leq n$, the causal policy $\policy_i$ follows the control strategy of $\policy$ at every time $t<i$, and from $t=i$ on switches to the optimal policy $\optimalpolicy$ defined in \eqref{optpolicydeff}. Clearly, $\policy_0=\optimalpolicy$, and $\policy_n=\policy$. So, one only needs to find $\instantcost{t}{\policy_k}-\instantcost{t}{\policy_{k-1}}$, for $1 \leq k \leq n$, and $0 \leq t \leq n-1$. For this purpose, let
	\begin{eqnarray*}
		D_0 &=& A_0+B_0 \Lmatrix{\para_0},\\
		P_0 &=& Q+\Lmatrix{\para_0}' R \Lmatrix{\para_0},\\
		M &=& B_0'\Kmatrix{\para_0}B_0+R,
	\end{eqnarray*}
	and fixing $k$, define the matrices
	\begin{eqnarray*}
		\Delta_{k} &=& B_0 \left(L_{k} - \Lmatrix{\para_0}\right),\\
		K_k &=& \sum\limits_{j=n-k}^{\infty} {D_0'}^{j} P_0 D_0^{j}.
	\end{eqnarray*}
	Let $\left\{ x(t) \right\}_{t=0}^{n-1}, \left\{ y(t) \right\}_{t=0}^{n-1}$ be the the state trajectory under the policies $\policy_k, \policy_{k-1}$, respectively. So,  
	\begin{eqnarray*}
		x(t) &=& y(t), \:\:\:\: \text{     for    }\:\:\:\: 0 \leq t \leq k-1, \\
		x(k) &=& D_0 x(k-1) + w(k)+ z_{k-1}, \\
		y(k) &=& D_0 x(k-1) + w(k),\\
		x(t)-y(t) &=& D_0^{t-k} z_{k-1}, \:\:\:\: \text{     for    }\:\:\:\: t \geq k,
	\end{eqnarray*}
	where $z_{k-1}=\Delta_{k-1} x(k-1) + B_0 v(k-1)$. Therefore, $\instantcost{t}{\policy_k} = \instantcost{t}{\policy_{k-1}}$ holds for $t \leq k-2$, and for $t=k-1$ we have 
	\begin{eqnarray*}
		\instantcost{t}{\policy_{k-1}} &=& x(t)'P_0 x(t),\\
		\instantcost{t}{\policy_{k}} &=& x(t)'P_{k-1} x(t)+ v(t)'Rv(t)+ 2 x(t)'L_{k-1}'Rv(t),
	\end{eqnarray*}
	where 
	\begin{eqnarray*}
		P_{k-1}= Q+L_{k-1}' R L_{k-1}.
	\end{eqnarray*}
	Since for $t\geq k$ we have
	\begin{eqnarray*}
		\instantcost{t}{\policy_k}-\instantcost{t}{\policy_{k-1}} = x(t)'P_0 x(t)-y(t)'P_0 y(t),
	\end{eqnarray*}
	the following holds:
	\begin{eqnarray*}
		&& \sum\limits_{t=0}^{n-1} \left[\instantcost{t}{\policy_k}-\instantcost{t}{\policy_{k-1}}\right] = \instantcost{k-1}{\policy_{k}} - \instantcost{k-1}{\policy_{k-1}}\\
		&+& \sum\limits_{t=k}^{n-1} \left[2 y(t)' P_0 D_0^{t-k} z_{k-1} + z_{k-1}' {D_0'}^{t-k} P_0 D_0^{t-k} z_{k-1}\right] .
 	\end{eqnarray*}
	Substituting for $y(t)$, and rearranging the terms, the above expression leads to
	\begin{eqnarray*}
		&& \sum\limits_{t=0}^{n-1} \left[\instantcost{t}{\policy_k}-\instantcost{t}{\policy_{k-1}}\right] = x(k-1)' E_{k-1} x(k-1)\\
		&+& v(k-1)' F_{k-1} v(k-1) + 2 x(k-1)' G_{k-1} v(k-1) \\
		&+& 2 \sum\limits_{j=k}^{n-1} w(j)' \left( K_n - K_j \right) D_0^{j-k} z_{k-1},
	\end{eqnarray*}
	where using the Lyapunov equation
	\begin{eqnarray*}
		\Kmatrix{\para_0}-D_0 \Kmatrix{\para_0} D_0 = P_0,
	\end{eqnarray*}
	the matrices $E_{k-1},F_{k-1},G_{k-1}$ can be calculated as follows:
	\begin{eqnarray*}
		E_{k-1} &=& L_{k-1}' R L_{k-1} - \Lmatrix{\para_0}' R \Lmatrix{\para_0} \\
		&+& D_0' \left( K_n - K_k \right) \Delta_{k-1}
		+ \Delta_{k-1}' \left( K_n - K_k \right) D_0 \\
		&+& \Delta_{k-1}' \left( K_n - K_k \right) \Delta_{k-1}, \\
		F_{k-1} &=& R + B_0' \left( K_n - K_k \right) B_0, \\
		G_{k-1} &=& L_{k-1}' R + \left( D_0'+ \Delta_{k-1}' \right) \left( K_n - K_k \right) B_0.		
	\end{eqnarray*}
	Since $K_n = \Kmatrix{\para_0}$, the definition of $\Lmatrix{\para_0}$ in \eqref{ricatti1} yields
	\begin{eqnarray*}
		E_{k-1} &=& \left(\Lmatrix{\para_0}-L_{k-1}\right)' M \left(\Lmatrix{\para_0}-L_{k-1}\right) - H_{k-1}, 
	\end{eqnarray*}
	where 
	\begin{eqnarray*}
		H_{k-1} = D_0' K_k \Delta_{k-1} + \Delta_{k-1}' K_k D_0 + \Delta_{k-1}' K_k \Delta_{k-1}.
	\end{eqnarray*}
	Because
	\begin{eqnarray*}
		\regret{n}{\policy}= \sum\limits_{k=1}^{n} \sum\limits_{t=0}^{n-1} \left[\instantcost{t}{\policy_k}-\instantcost{t}{\policy_{k-1}}\right],
	\end{eqnarray*}
	the summation of the above expressions leads to
	\begin{eqnarray} \label{regretdecompos1}
	\regret{n}{\policy} = \phi_n+ \zeta_n + \xi_n + \psi_n + \sum\limits_{j=1}^{n-1} s(j)' w(j),
	\end{eqnarray}	 
	where $s(j) = 2 \sum\limits_{k=1}^{j} \left(K_n - K_j \right) D_0^{j-k} z_{k-1}$, and 
	\begin{eqnarray}
	\phi_n &=& -\sum\limits_{k=0}^{n-1} x(k)'H_{k}x(k), \label{phideff}\\
	\zeta_n &=& \sum\limits_{k=0}^{n-1} v(k)'F_{k}v(k), \notag \\
	\xi_n &=& \sum\limits_{k=0}^{n-1} x(k)'G_{k}v(k), \notag \\
	\psi_n &=& \sum\limits_{k=0}^{n-1} \norm{M^{1/2}\left(L_{k}-\Lmatrix{\para_0}\right) x(k)}{2}^2 . \notag
	\end{eqnarray}
	
	According to $K_n = \Kmatrix{\para_0}$, positive semi-definiteness of $K_k$ implies
	\begin{eqnarray} \label{boundforvv}
	\zeta_n \leq \eigmax{M} \totalbound{v}.
	\end{eqnarray}
	Further, one can bound $\xi_n$:
	\begin{eqnarray*}
		\left|x(k)'G_{k}v(k)\right| \leq \ubound{g_k} \left( \ubound{\ell_k} +  \ubound{x_k} \right) \ubound{v_k},
	\end{eqnarray*}
	where
	\begin{eqnarray*}
		\ubound{g_k} &=& \left( 1 \vee \Mnorm{\Lmatrix{\para_0}}{2} \right) \Mnorm{R}{2} \\
		&+& \left( \Mnorm{D_0}{2} \vee \Mnorm{B_0}{2} \right) \Mnorm{K_n-K_{k+1}}{2} \Mnorm{B_0}{2}. 
	\end{eqnarray*} 
	So, applying Azuma's Inequality \cite{tropp2012user} we obtain
	\begin{eqnarray} \label{boundforxv}
	\PP{ \frac{\xi_n^2}{8 \log \left(\frac{10 n^2}{\delta}\right)} > \sum\limits_{k=0}^{n-1} \ubound{g_k}^2 \left( \ubound{\ell_k} +  \ubound{x_k} \right)^2 \ubound{v_k}^2 } \leq \frac{\delta}{5 n^2}.
	\end{eqnarray}
	To proceed, note that $\left| s(j)'w(j) \right| \leq \ubound{s_j} \: \ubound{w_j} $, where
	\begin{eqnarray*}
		\ubound{s_j} = 2\Mnorm{K_n - K_j}{2} \Mnorm{B_0}{2}  \sum\limits_{k=1}^{j} \Mnorm{D_0^{j-k}}{2} \left(  \ubound{\ell_{k-1}}+ \ubound{v_{k-1}} \right).
	\end{eqnarray*}
	Therefore, Azuma's Inequality \cite{tropp2012user} implies that with probability at least $1- 0.2 \delta n^{-2}$, we have
	\begin{eqnarray} \label{boundforsw} 
		\left(\sum\limits_{j=1}^{n-1} s(j)'w(j)\right)^2 \leq 8 \log \left(\frac{10 n^2}{\delta}\right)\sum\limits_{j=1}^{n-1} \ubound{s_j}^2 \ubound{w_j}^2 . 
	\end{eqnarray}
	Next, in order to investigate $\phi_n$, since
	\begin{eqnarray*}
		x(k) &=& \left( D_0+\Delta_{k-1} \right) x(k-1)+ B_0v(k-1)+w(k), \\
		K_k &=& D_0 K_{k+1} D_0, 
	\end{eqnarray*}
	the following holds:
	\begin{eqnarray*}
		&&x(k)' D_0' K_{k+1} D_0 x(k) \\
		&=& x(k-1)' \left(D_{0}+\Delta_{k-1}\right)' K_{k} \left(D_{0}+\Delta_{k-1}\right) x(k-1) \\
		&+& \surnoise{k}' K_{k} \surnoise{k} + 2 \surnoise{k}' K_{k} \left(D_{0}+\Delta_{k-1}\right) x(k-1),
	\end{eqnarray*}
	where $\surnoise{k}= B_0v(k-1)+w(k)$, for $k \geq 1$. Substitute the above expression in~\eqref{phideff}, denote $\surnoise{0}=x(0)$, and write 
	\begin{eqnarray*}
		H_k= \left( D_0+\Delta_{k} \right)' K_{k+1} \left( D_0+\Delta_{k} \right) - D_0'K_{k+1} D_0,
	\end{eqnarray*}
	to obtain: 
	\begin{eqnarray*} 
		\phi_n &=& -x(n)' \Kmatrix{\para_0} x(n) + \sum\limits_{k=0}^{n} \surnoise{k}' K_{k} \surnoise{k} \\
		&+& \sum\limits_{k=0}^{n-1} 2 \surnoise{k+1}' K_{k+1} \left(D_{0}+\Delta_{k}\right) x(k).
	\end{eqnarray*}
	So, defining $\optstate{n}=\sum\limits_{j=0}^n D_0^{n-j} \surnoise{j}$, and using  
	\begin{eqnarray*}
		\left( D_0 + \Delta_{k} \right) x(k)= \sum\limits_{j=0}^k D_0^{k-j} \left( D_0 \surnoise{j} + \Delta_{j} x(j) \right),
	\end{eqnarray*}
	we get 
	\begin{eqnarray*}
		\phi_n &=& \optstate{n}' K_n \optstate{n} - x(n)' K_n x(n) + \sum\limits_{k=1}^{n} \surnoise{k}' h(k) ,
	\end{eqnarray*}
	where $h(k)=2 K_{k} \sum\limits_{t=0}^{k-1} D_0^{k-t-1} \Delta_{t} x(t)$. Letting 
	\begin{eqnarray*}
		\ubound{b_k} &=& \Mnorm{B_0}{2} \ubound{v_{k-1}} + \ubound{w_k},\\
		\ubound{h_k} &=& 2 \Mnorm{K_k}{2} \Mnorm{B_0}{2} \sum\limits_{t=0}^{k-1} \Mnorm{D_0^{k-t-1}}{2} \ubound{\ell_t},
	\end{eqnarray*}
	according to Azuma's Inequality~\cite{tropp2012user}, $\left|\surnoise{k}' h(k) \right| \leq \ubound{b_k} \: \ubound{h_k}$ implies that with probability at least $1-0.2 \delta n^{-2}$, the following holds:
	\begin{eqnarray} \label{boundforxx1}
		\left(\sum\limits_{k=1}^{n} \surnoise{k}' h(k)\right)^2 \leq 8 \log \left( \frac{10 n^2}{\delta}\right) \sum\limits_{k=1}^n \ubound{b_k}^2 \ubound{h_k}^2 .
	\end{eqnarray}
	Moreover, we have $x(n) - \optstate{n} = \sum\limits_{j=0}^{n-1} D_0^{n-j} \Delta_{j} x(j)$, as well as
	\begin{eqnarray} 
		&& \frac{\optstate{n}' K_n \optstate{n} - x(n)' K_n x(n)}{\Mnorm{K_n}{2}} \notag \\
		&\leq& \norm{\optstate{n} - x(n)}{2}^2 + 2 \norm{\optstate{n} - x(n)}{2} \norm{\optstate{n}}{2}. \label{boundforxx2}
	\end{eqnarray}
	
	Then, stability of $D_0$ clearly implies that all the following quantities are finite, with upper bounds depending only on $\para_0,Q,R$:  
	\begin{eqnarray*}
		&& \sum\limits_{k=0}^\infty \Mnorm{D_0^k}{2},\\
		&& \sup\limits_{0 \leq k < n} \Mnorm{K_k}{2},\\
		&& \sup\limits_{0 \leq k < n} \Mnorm{K_n - K_k}{2},\\
		&& \sup\limits_{0 \leq k < n} \ubound{g_k}, \\
		&& \sup\limits_{n \geq 1} \frac{\norm{\optstate{n}}{2}}{\ubound{w^*}+\totalbound{v}^{1/2}}, \\
		&& \sup\limits_{n \geq 1} \frac{\norm{\optstate{n}-x(n)}{2}}{\totalbound{\ell}^{1/2}}, \\
		&&\sup\limits_{n \geq 1} \frac{\sum\limits_{k=1}^{n} \ubound{b_k}^2 \ubound{h_k}^2}{\ubound{w^*}^2 \sum\limits_{t=0}^{n-1} \left(\ubound{\ell_t}+\ubound{v_t}\right)^2}, \\
		&& \sup\limits_{n \geq 1} \frac{\sum\limits_{j=1}^{n-1} \ubound{s_j}^2 \ubound{w_j}^2}{\ubound{w^*}^2 \sum\limits_{t=0}^{n-1} \left(\ubound{\ell_t}+\ubound{v_t}\right)^2}.
	\end{eqnarray*}
	Using the above bounded quantities, we can specify the constant $\ssconstant_1$ in Theorem~\ref{RegretTheorem} as follows. Indeed, since $\psi_n \leq \eigmax{M} \totalbound{\ell}$, plugging \eqref{boundforvv}, \eqref{boundforxv}, \eqref{boundforsw}, \eqref{boundforxx1}, and \eqref{boundforxx2} in \eqref{regretdecompos1}, with probability at least $1-0.6 \delta n^{-2}$ it holds that
	\begin{eqnarray*} 
		\frac{\regret{n}{\policy}}{\ssconstant_1} \leq \totalbound{v}+ \totalbound{\ell} + \totalbound{\policy} \log^{1/2} \left( n \delta^{-1} \right) ,
	\end{eqnarray*}
	where $\ssconstant_1 < \infty$ is fixed. Taking a union bound, we get the desired result of Theorem \ref{RegretTheorem} since $\sum\limits_{n=1}^\infty 0.6n^{-2}<1$. Moreover, since
	\begin{eqnarray*}
		\E{ \xi_n } &=& 0,\\
		\E{ s(j)'w(j)} &=& 0,\\
		\E{\surnoise{k}'h(k)} &=& 0,\\
		\sup\limits_{n \geq 1} \E{\norm{\optstate{n}}{2}^2 } &\leq& \sum\limits_{j=0}^\infty \Mnorm{D_0^{j}}{2}^2 \E{ \norm{\surnoise{j}}{2}^2  } < \infty,
	\end{eqnarray*}
	clearly Corollary~\ref{ExpRegretCor} is concluded from \eqref{phideff}, \eqref{boundforvv}, $x(n)'K_n x(n) \geq 0$, and $\psi_n \leq \eigmax{M} \totalbound{\ell}$.
\subsection{\bf Proof of Theorem \ref{EstTheorem}}
	First, solving for the least squares estimate in \eqref{LSE}, we get
	\begin{eqnarray*}
		\estpara{n} \empiricalcovmat{} = \sum\limits_{t=0}^{n-1} x(t+1) \left[x(t)' , u(t)'\right], 
	\end{eqnarray*}
	where 
	\begin{eqnarray} \label{empiricalcovdeff}
	\empiricalcovmat{}=\sum\limits_{t=0}^{n-1} \begin{bmatrix}
	x(t) \\ u(t)
	\end{bmatrix} \left[x(t)' , u(t)'\right],
	\end{eqnarray} 
	is the (unnormalized) empirical covariance matrix of the covariates $x(t),u(t)$. Since $x(t+1)=\para_0 \begin{bmatrix}
	x(t) \\ u(t)
	\end{bmatrix} + w(t+1)$, we obtain
	\begin{eqnarray} \label{normaleq}
		\left(\estpara{n} - \para_0 \right) \empiricalcovmat{} = \sum\limits_{t=0}^{n-1} w(t+1) \left[x(t)' , u(t)'\right]. 
	\end{eqnarray}
	In the sequel we investigate $\empiricalcovmat{}$. The design of the perturbed input according to $u(t)=L_t x(t)+ v(t)$ leads to the closed-loop dynamics 
	\begin{eqnarray*}
		x(t+1)=D_t x(t)+z(t),
	\end{eqnarray*}
	where 
	\begin{eqnarray*}
	D_t &=& A_0+B_0L_t, \\
	z(t) &=& B_0v(t)+w(t+1). 
	\end{eqnarray*}
	So, in order to study the $p \times p$ left upper block of $\empiricalcovmat{}$ which is the Gram matrix of the output signal, we write
	\begin{eqnarray*}
		\empiricalcovmat{x} &=& \sum\limits_{t=0}^{n-1} x(t)x(t)' = x(0)x(0)' + \sum\limits_{t=0}^{n-2} x(t+1)x(t+1)' \\
		&=& \covmat{z} +x(0)x(0)' +\sum\limits_{t=0}^{n-2} \left[ M_t+D_tx(t)x(t)'D_t'\right],
	\end{eqnarray*}
	where $\covmat{z}=\sum\limits_{t=0}^{n-2} \E{z(t)z(t)'}$, and
	\begin{eqnarray*}
		M_t = z(t)z(t)' - \E{z(t)z(t)'} + D_tx(t)z(t)' + z(t)x(t)'D_t'.
	\end{eqnarray*}
	Then, $M_t$ is a martingale difference sequence that according to Definition \ref{mineigdefn} is bounded: 
	\begin{eqnarray*}
	\eigmax{M_t} \leq \ubound{y_t} \left( \Mnorm{B_0}{2}\ubound{v_t} +\ubound{w_{t+1}}\right). 
	\end{eqnarray*}
	So, using $\totalbound{1}$ in Definition \ref{mineigdefn}, Azuma's Matrix Inequality \cite{tropp2012user} implies that 	
	\begin{eqnarray*}
		&& \PP{\eigmax{\sum\limits_{t=0}^{n-2} M_t} > 2^{-1/2}\eigmin{\covmat{z}} } \\
		&\leq& 2p \: \exp \left(16^{-1} \totalbound{1}^{-2} \eigmin{\covmat{z}}^2 \right) \leq 4^{-1}\delta,
	\end{eqnarray*}
	where in the last inequality above we used \eqref{minSSforw} and $\eigmin{\covmat{z}} \geq \eigmin{\covmat{w}}$. Therefore, positive semidefiniteness of $x(0)x(0)' +\sum\limits_{t=0}^{n-2} D_tx(t)x(t)'D_t'$ leads to
	\begin{eqnarray} \label{empiricalmineig1}
		\PP{\eigmin{\empiricalcovmat{x}}< 0.29 \eigmin{\covmat{w}}} \leq 4^{-1} \delta.
	\end{eqnarray}
	To proceed, note that letting 
	\begin{eqnarray*}
		\tilde{L}_t = \begin{bmatrix}
			I_p \\ L_t
		\end{bmatrix}, \:\:\:
		\tilde{v}(t) = \begin{bmatrix}
			0_p \\ v(t)
		\end{bmatrix}, 
	\end{eqnarray*}
	we have
	\begin{eqnarray*}
		\empiricalcovmat{}= \covmat{\tilde{v}}+\sum\limits_{t=0}^{n-1} \left(\tilde{L}_t x(t)x(t)' \tilde{L}_t' + N_t\right),
	\end{eqnarray*}
	where $\covmat{\tilde{v}}=\sum\limits_{t=0}^{n-1} \E{\tilde{v}(t)\tilde{v}(t)'}$, and
	\begin{eqnarray*}
	N_t=\tilde{L}_t x(t)\tilde{v}(t)' + \tilde{v}(t)x(t)' \tilde{L}_t'+ \tilde{v}(t)\tilde{v}(t)' - \E{\tilde{v}(t)\tilde{v}(t)'}.
	\end{eqnarray*}
	We show that
	\begin{eqnarray} \label{empiricalmineig2}
		\PP{ \eigmax{\sum\limits_{t=0}^{n-1} N_t} \geq 2^{-1/2} \eigmin{\covmat{v}}} \leq 4^{-1} \delta. 
	\end{eqnarray}
	For this purpose, we leverage Azuma's Matrix Inequality \cite{tropp2012user} as follows. The martingale difference sequence $N_t$ is bounded:
	\begin{eqnarray*}
		\eigmax{N_t} \leq \ubound{v_t} \left( 2 \left[ 1+\Mnorm{\Lmatrix{\para_0}}{2} \right] \ubound{x_t} + 2\ubound{\ell_t} + \ubound{v_t} \right).
	\end{eqnarray*}
	According to Definition \ref{mineigdefn}, \eqref{minSSforv} leads to \eqref{empiricalmineig2}. Next, note that the upper $p \times p$ block of $\tilde{L}_t$ is the identity matrix $I_p$, and the upper $p \times 1$ block of $\tilde{v}(t)$ is the zero vector $0_p$. Hence, putting \eqref{empiricalmineig1}, \eqref{empiricalmineig2} together, we get the following for the quantity $\lbound{\lambda_n}$ defined in Definition \ref{mineigdefn}:
	\begin{eqnarray} \label{empiricalmineig3}
	\PP{ \eigmin{\empiricalcovmat{}} < 0.29 \: \lbound{\lambda_n} } \leq 2^{-1} \delta. 
	\end{eqnarray}
	Going back to \eqref{normaleq}, we use the following non-asymptotic result for matrix valued martingales \cite{abbasi2011improved}. The asymptotic version can be found in \cite{lai1982least}.
	\begin{lemm} \cite{abbasi2011improved} \label{YasinLemma}
		With probability at least $1-2^{-1}\delta$, the following holds for all $n \geq 1$:
		\begin{eqnarray*}
			&& \Mnorm{\left(\lbound{\lambda_n}I_q+ \empiricalcovmat{}\right)^{-1/2} \sum\limits_{t=0}^{n-1} \begin{bmatrix}
					x(t) \\ u(t)
			\end{bmatrix} \frac{w(t+1)'}{\ubound{w_{t+1}}} }{2}^2 \\
			&\leq& pq \log \eigmax{\lbound{\lambda_n}I_q+ \empiricalcovmat{}} \\
			&-& pq \log \lbound{\lambda_n} +2 p \log \left( 2p \delta^{-1} \right).
		\end{eqnarray*}
	\end{lemm}  
	Then, we clearly have $\max\limits_{1 \leq t \leq n} \ubound{w_t} \leq \ubound{w^*}$, as well as
	\begin{eqnarray*}
		\eigmax{\empiricalcovmat{}} &\leq& \sum\limits_{t=0}^{n-1} \left(\norm{x(t)}{2}^2 + \norm{u(t)}{2}^2\right) \\
		&\leq& \left( 1 + 2 \Mnorm{\Lmatrix{\para_0}}{2}^2 \right) \totalbound{x} + 2 \totalbound{\ell} + 2 \totalbound{v}.
	\end{eqnarray*}
	Moreover, for an arbitrary $\para \in \R^{p \times q}$, we have 
	\begin{eqnarray*}
		\Mnorm{\para \empiricalcovmat{}^{-1} \para'}{2} \leq 1.29 \Mnorm{\para \left( \lbound{\lambda_n}I_q + \empiricalcovmat{}\right)^{-1} \para'}{2},
	\end{eqnarray*}
	as long as $\eigmin{\empiricalcovmat{}} \geq 0.29 \lbound{\lambda_n}$. Therefore, plugging \eqref{empiricalmineig3} and the result of Lemma \ref{YasinLemma} in equation \eqref{normaleq}, we have
	\begin{eqnarray*}
		&&0.29 \lbound{\lambda_n}\Mnorm{\estpara{n}-\para_0}{2}^2 \\ &\leq& \Mnorm{\left( \estpara{n}-\para_0 \right) \empiricalcovmat{} \left( \estpara{n}-\para_0 \right)'}{2} \\
		&\leq& 1.29 \Mnorm{ \left( \lbound{\lambda_n}I_q + \empiricalcovmat{}\right)^{-1/2} \left(\estpara{n}-\para_0\right)'}{2}^2 \\
		&\leq& 2pq \ubound{w^*}^2 \log \left[ 4p \left( \left( 1 + \Mnorm{\Lmatrix{\para_0}}{2} \right)^2 \totalbound{x} + \totalbound{\ell} + \totalbound{v} \right) \delta^{-1}\right],
	\end{eqnarray*}
	with probability at least $1-\delta$, which is the desired result.
	
\subsection{\bf Proof of Theorem \ref{GreedyRates}}
	First, suppose that $\max\limits_{1 \leq t \leq n} \norm{w(t)}{2} \leq \ubound{w^*}$. To find the growth rate of the regret, according to Theorem \ref{RegretTheorem} it suffices to determine $\totalbound{\ell},\totalbound{v},\totalbound{\adaptivepolicy}$. Further, by \eqref{LipschitzL}, Theorem \ref{EstTheorem} implies that in order to determine the above quantities, one needs to address the behavior of $\totalbound{1},\totalbound{2},\lbound{\lambda_n},\totalbound{x},$. 
	
	Let $\lbound{\lambda_n}$ be as defined in Definition \ref{mineigdefn}. At the end of epoch $m$ that corresponds to the time step $n = \lfloor \rrate^m \rfloor$, the adaptive policy $\adaptivepolicy$ updates the parameter estimates according to \eqref{LSE}. Then, the second part of Assumption \ref{tailcondition} and the design of perturbation in \eqref{PerturbDesign}, lead to the following lower bound:
	\begin{eqnarray} \label{Greedymineig}
	\lambda_n \geq \left( \sigma_0 \wedge \lbound{C} \left( \rrate - 1 \right) \log^{-2} \rrate \right) n^{1/2} \log^{2} n.
	\end{eqnarray} 
	Since the adaptive policy $\adaptivepolicy$ is stabilizing the system, we have $\ubound{x_t} \leq \eta_x \ubound{w^*}$, $\ubound{\ell_t} \leq \eta_\ell \ubound{w^*}$, for some constants $\eta_x,\eta_\ell< \infty$ \cite{faradonbeh2017finite}. Thus, since the perturbation signal $v(t)$ is diminishing with the rate specified in \eqref{PerturbDesign}, we have
	\begin{eqnarray*}
		\totalbound{1} &\leq& \eta_1 \ubound{w^*}^2 n^{1/2}, \\ \totalbound{2} &\leq& \eta_2 \ubound{w^*} n^{1/4} \log n, 
	\end{eqnarray*}
	for some $\eta_1,\eta_2 < \infty$. Therefore, according to \eqref{Greedymineig}, both \eqref{minSSforw}, \eqref{minSSforv} will be satisfied if 
	\begin{eqnarray} \label{minSSGreedy}
		- \eta_0 \log \delta \leq \ubound{w^*}^{-2} n^{1/2} \log^{2} n \vee \ubound{w^*}^{-4} n,
	\end{eqnarray}
	where $\eta_0< \infty$ is a fixed constant determined by $\lbound{C},\ubound{C},\sigma_0, \eta_1,\eta_2 , \log q$. Let $n_0$ be large enough such that \eqref{minSSGreedy} holds for all $n \geq n_0$. 
	
	Next, plugging \eqref{Greedymineig} in Theorem \ref{EstTheorem}, we obtain 
	\begin{eqnarray} \label{GreedyLearningRate}
		\sup\limits_{n \geq n_0} \frac{n^{1/2} \log^{2} n \Mnorm{\estpara{n}-\para_0}{2}^2}{\ubound{w^*}^2 \log \left( n \ubound{w^*}^2 \delta^{-1} \right)} < \tilde{\ssconstant_4},
	\end{eqnarray}
	with probability at least $1-\delta$. In addition, using \eqref{LipschitzL}, we get the following high probability result:
	\begin{eqnarray} \label{GreedyGainRate}
	\sup\limits_{n \geq n_0} \frac{ \left( n \log^{4} n \right)^{1/4}\Mnorm{\Lmatrix{\para_0}-L_n}{2}}{ \ubound{w^*} \log^{1/2} \left( n \ubound{w^*}^2 \delta^{-1} \right)} < \infty.
	\end{eqnarray}
	By Definition \ref{signalbounds}, \eqref{GreedyGainRate} yields 
	\begin{eqnarray}
		\sup\limits_{n \geq n_0} \frac{\totalbound{\ell}}{\left(n\log^{-2} n\right)^{1/2}  \ubound{w^*}^4 \log \left( n \ubound{w^*}^2 \delta^{-1} \right) } &\leq& \ssconstant_\ell , \label{GreedyPlanningRate1}\\
		\sup\limits_{n \geq n_0} \frac{\totalbound{\adaptivepolicy}}{\ubound{w^*} \left( \totalbound{\ell} + \totalbound{v} \right)^{1/2}} &\leq& \ssconstant_{\adaptivepolicy} , \label{GreedyPlanningRat2}
	\end{eqnarray}
	where $\ssconstant_\ell, \ssconstant_{\adaptivepolicy}$ are fixed and finite. 
	
	Now, according to Assumption \ref{tailcondition}, with probability at least $1-\delta$ we have \cite{faradonbeh2017finite}:
	\begin{eqnarray} \label{weibulltail}
		\max\limits_{1 \leq t \leq n} \norm{w(t)}{2} \leq \tailcoeff^{1/\tailexp} \log^{1/\tailexp} \left( \tailconst n \delta^{-1} \right).
	\end{eqnarray}
	Thus, substituting the above value for $\ubound{w^*}$ in \eqref{GreedyLearningRate}, we get the desired result for the identification error. Moreover, since
	\begin{eqnarray*}
		{\totalbound{v}} \leq \rrate \log^{-2} \rrate  \ubound{\ssconstant} n^{1/2} \log^{2} n ,
	\end{eqnarray*}
	\eqref{GreedyPlanningRate1}, \eqref{GreedyPlanningRat2}, and Theorem \ref{RegretTheorem} imply the desired result for the regret bound. Finally, for a fixed constant $\ssconstant_0$, letting 
	\begin{eqnarray} \label{n_0deff}
		n_0 \geq \ssconstant_0 \left(\log^{2+4/\tailexp} \delta^{-1}\right) \left(\log\log \delta^{-1}\right)
	\end{eqnarray}
	is sufficient to satisfy \eqref{minSSGreedy}.

\subsection{\bf Proof of Theorem \ref{IdGreedyRates}}
	First, similar to the proof of Theorem \ref{GreedyRates}, assume $\max\limits_{1 \leq t \leq n} \norm{w(t)}{2} \leq \ubound{w^*}$. Let $\optpara{\lfloor \rrate^i \rfloor}$ be the least squares estimate over all parameter space $\R^{p \times q}$ given by \eqref{LSE} for the input-output observations being collected until the end of epoch $i$. Note that $\optpara{\lfloor \rrate^i \rfloor}$ is not necessarily equal to the solution of \eqref{IDLSE} denoted by $\estpara{\lfloor \rrate^i \rfloor}$. Letting $n=\lfloor \rrate^i \rfloor$, define the Gram matrix $\empiricalcovmat{}$ according to \eqref{empiricalcovdeff}. Then, $\para_0 \in \tempparaspace{0}$ implies that
	\begin{eqnarray} 
	&& \tr{ \left(\estpara{\lfloor \rrate^i \rfloor} - \para_0 \right) \empiricalcovmat{} \left(\estpara{\lfloor \rrate^i \rfloor} - \para_0 \right)' } \notag \\
	&\leq& 4 \tr{ \left(\optpara{\lfloor \rrate^i \rfloor} - \para_0 \right) \empiricalcovmat{} \left(\optpara{\lfloor \rrate^i \rfloor} - \para_0 \right)'}.\label{lossbound1}
	\end{eqnarray}
	Then, using \eqref{normaleq}, Lemma \ref{YasinLemma} leads to
	\begin{eqnarray} 
	&&\Mnorm{ \empiricalcovmat{}^{1/2} \left(\optpara{\lfloor \rrate^i \rfloor} - \para_0 \right)'}{2}^2 \notag \\
	&& \leq pq \ubound{w^*}^2 \log \frac{2p \eigmax{\empiricalcovmat{}}}{\delta} , \label{lossbound2}
	\end{eqnarray}
	with probability at least $1-\delta/2$. Next, note that during each epoch the parameter estimates and so the feedback gains are fixed. So, let $L_i$ be the feedback gain during the $i$-th epoch: $L_i=\Lmatrix{\estpara{\lfloor \rrate^i \rfloor}}$, and define $\tilde{L}_i= \left[I_p, L_i'\right]'$. Hence, the stable closed-loop dynamics takes the form
	\begin{eqnarray*}
		x(t+1)=D_i x(t)+B_0v(t)+w(t+1), 
	\end{eqnarray*}
	where $D_i=\para_0 \tilde{L}_i$. Further, define
	\begin{eqnarray*}
		V_i= \sum\limits_{t=\lfloor \rrate^{i-1} \rfloor }^{\lfloor \rrate^i \rfloor -1}x(t)x(t)',\:\:\: U_i = \tilde{L}_{i} V_{i} \tilde{L}_i' + W_i,
	\end{eqnarray*}
	where $\tilde{v}(t)=\left[0_p,v(t)'\right]'$, and
	\begin{eqnarray*}
		W_i=\sum\limits_{t=\lfloor \rrate^{i-1} \rfloor}^{\lfloor \rrate^{i} \rfloor-1}  \left[\tilde{L}_i x(t)\tilde{v}(t)' +  \tilde{v}(t)x(t)' \tilde{L}_i' + \tilde{v}(t) \tilde{v}(t)'\right].
	\end{eqnarray*}
	Thus, \eqref{lossbound1}, \eqref{lossbound2} yield
	\begin{eqnarray*}
		\Mnorm{U_i^{1/2} \left( \estpara{\lfloor \rrate^i \rfloor} - \para_0 \right) }{2}^2 \leq 
		4 p^2 q \ubound{w^*}^2 \log \left(2p \eigmax{\empiricalcovmat{}}\delta^{-1}\right) .
	\end{eqnarray*}
	Hence, applying Cauchy-Schwarz inequality, and using the design of perturbation; $\rrate^{i} \ubound{v_i}^2 < \ubound{\ssconstant}$, we obtain
	\begin{eqnarray*}
		\Mnorm{V_i^{1/2} \tilde{L}_{i}' \left( \estpara{\lfloor \rrate^i \rfloor} - \para_0 \right) }{2}^2 - \ubound{\ssconstant} \Mnorm{V_i^{1/2} \tilde{L}_{i}' \left( \estpara{\lfloor \rrate^i \rfloor} - \para_0 \right) }{2} \\
		\leq 
		4 p^2 q \ubound{w^*}^2 \log \left(2p \eigmax{\empiricalcovmat{}}\delta^{-1}\right) 
	\end{eqnarray*}
	Then, we use the following result.
	\begin{lemm} \cite{faradonbeh2017finite} \label{covmatbounds}
		With probability at least $1-\delta/2$ we have:
		\begin{eqnarray*}
			\frac{\sigma_0}{2} \leq \inf\limits_{i \geq m_0} \frac{\eigmin{V_i}}{\rrate^{i}} \leq \sup\limits_{i \geq m_0} \frac{\eigmax{V_i}}{\rrate^{i}} \leq \eta_0 \ubound{w^*}^2,
		\end{eqnarray*} 
		for a constant $\eta_0<\infty$, and $m_0 = \log \left( \eta_0 \sigma_0^{-1} \ubound{w^*} \log \delta^{-1} \right)$.
	\end{lemm} 
	By Lemma \ref{covmatbounds}, and \eqref{Sideq} we get
	\begin{eqnarray*}
		\sup\limits_{n \geq \rrate^{m_0}} \frac{\Mnorm{\Lmatrix{\estpara{n}} - \Lmatrix{\para_0} }{2}^2 }{ n^{-1} \ubound{w^*}^2 \log \left( n \delta^{-1} \right) } \leq \tilde{\ssconstant} < \infty, 
	\end{eqnarray*}
	with probability at least $1-\delta$. Finally, according to \eqref{weibulltail}, Theorem \ref{RegretTheorem} implies the desired result.
\newpage

\fi


\begin{thebibliography}{10}
\providecommand{\url}[1]{#1}
\csname url@samestyle\endcsname
\providecommand{\newblock}{\relax}
\providecommand{\bibinfo}[2]{#2}
\providecommand{\BIBentrySTDinterwordspacing}{\spaceskip=0pt\relax}
\providecommand{\BIBentryALTinterwordstretchfactor}{4}
\providecommand{\BIBentryALTinterwordspacing}{\spaceskip=\fontdimen2\font plus
\BIBentryALTinterwordstretchfactor\fontdimen3\font minus
  \fontdimen4\font\relax}
\providecommand{\BIBforeignlanguage}[2]{{%
\expandafter\ifx\csname l@#1\endcsname\relax
\typeout{** WARNING: IEEEtran.bst: No hyphenation pattern has been}%
\typeout{** loaded for the language `#1'. Using the pattern for}%
\typeout{** the default language instead.}%
\else
\language=\csname l@#1\endcsname
\fi
#2}}
\providecommand{\BIBdecl}{\relax}
\BIBdecl

\bibitem{kailath1980linear}
T.~Kailath, \emph{Linear systems}.\hskip 1em plus 0.5em minus 0.4em\relax
  Prentice-Hall Englewood Cliffs, NJ, 1980, vol. 156.

\bibitem{meyn2008control}
S.~Meyn, \emph{Control techniques for complex networks}.\hskip 1em plus 0.5em
  minus 0.4em\relax Cambridge University Press, 2008.

\bibitem{faradonbeh2016optimality}
M.~K.~S. Faradonbeh, A.~Tewari, and G.~Michailidis, ``Optimality of fast
  matching algorithms for random networks with applications to structural
  controllability,'' \emph{IEEE Transactions on Control of Network Systems},
  vol.~4, no.~4, pp. 770--780, 2017.

\bibitem{marino1995nonlinear}
R.~Marino and P.~Tomei, \emph{Nonlinear control design: geometric, adaptive and
  robust}.\hskip 1em plus 0.5em minus 0.4em\relax Prentice Hall London, 1995,
  vol.~1.

\bibitem{li2013stabilization}
C.~Li and J.~Lam, ``Stabilization of discrete-time nonlinear uncertain systems
  by feedback based on ls algorithm,'' \emph{SIAM Journal on Control and
  Optimization}, vol.~51, no.~2, pp. 1128--1151, 2013.

\bibitem{lazic2018data}
N.~Lazic, T.~Lu, C.~Boutilier, M.~Ryu, E.~Wong, B.~Roy, and G.~Imwalle, ``Data
  center cooling using model-predictive control.''\hskip 1em plus 0.5em minus
  0.4em\relax NIPS, 2018.

\bibitem{pesaran2005small}
M.~H. Pesaran and A.~Timmermann, ``Small sample properties of forecasts from
  autoregressive models under structural breaks,'' \emph{Journal of
  Econometrics}, vol. 129, no. 1-2, pp. 183--217, 2005.

\bibitem{faradonbeh2018finite}
M.~K.~S. Faradonbeh, A.~Tewari, and G.~Michailidis, ``Finite time
  identification in unstable linear systems,'' \emph{Automatica}, vol.~96, pp.
  342--353, 2018.

\bibitem{bertsekas1995dynamic}
D.~P. Bertsekas, \emph{Dynamic programming and optimal control}.\hskip 1em plus
  0.5em minus 0.4em\relax Athena Scientific Belmont, MA, 1995, vol.~1, no.~2.

\bibitem{dorato1995linear}
P.~Dorato, C.~T. Abdallah, V.~Cerone, and D.~H. Jacobson,
  \emph{Linear-quadratic control: an introduction}.\hskip 1em plus 0.5em minus
  0.4em\relax Prentice Hall Englewood Cliffs, NJ, 1995.

\bibitem{guo1991aastrom}
L.~Guo and H.-F. Chen, ``The {\aa}strom-wittenmark self-tuning regulator
  revisited and els-based adaptive trackers,'' \emph{IEEE Transactions on
  Automatic Control}, vol.~36, no.~7, pp. 802--812, 1991.

\bibitem{lai1985asymptotically}
T.~L. Lai and H.~Robbins, ``Asymptotically efficient adaptive allocation
  rules,'' \emph{Advances in applied mathematics}, vol.~6, no.~1, pp. 4--22,
  1985.

\bibitem{campi1998adaptive}
M.~C. Campi and P.~Kumar, ``Adaptive linear quadratic gaussian control: the
  cost-biased approach revisited,'' \emph{SIAM Journal on Control and
  Optimization}, vol.~36, no.~6, pp. 1890--1907, 1998.

\bibitem{bittanti2006adaptive}
S.~Bittanti and M.~C. Campi, ``Adaptive control of linear time invariant
  systems: the “bet on the best” principle,'' \emph{Communications in
  Information \& Systems}, vol.~6, no.~4, pp. 299--320, 2006.

\bibitem{abbasi2011regret}
Y.~Abbasi-Yadkori and C.~Szepesv{\'a}ri, ``Regret bounds for the adaptive
  control of linear quadratic systems.'' in \emph{COLT}, 2011, pp. 1--26.

\bibitem{ibrahimi2012efficient}
M.~Ibrahimi, A.~Javanmard, and B.~V. Roy, ``Efficient reinforcement learning
  for high dimensional linear quadratic systems,'' in \emph{Advances in Neural
  Information Processing Systems}, 2012, pp. 2636--2644.

\bibitem{faradonbeh2017finite}
M.~K.~S. Faradonbeh, A.~Tewari, and G.~Michailidis, ``Optimism-based adaptive
  regulation of linear-quadratic systems,'' \emph{IEEE
  	Transactions on Automatic Control}, available online: arXiv:1711.07230.

\bibitem{faradonbeh2018optimality}
------, ``On optimality of adaptive linear-quadratic regulators,'' \emph{arXiv
  preprint arXiv:1806.10749}, 2018.

\bibitem{imer2006optimal}
O.~C. Imer, S.~Y{\"u}ksel, and T.~Ba{\c{s}}ar, ``Optimal control of {LTI}
  systems over unreliable communication links,'' \emph{Automatica}, vol.~42,
  no.~9, pp. 1429--1439, 2006.

\bibitem{ouyang2018optimal}
Y.~Ouyang, S.~M. Asghari, and A.~Nayyar, ``Optimal infinite horizon
  decentralized networked controllers with unreliable communication,''
  \emph{arXiv preprint arXiv:1806.06497}, 2018.

\bibitem{abbasi2011improved}
Y.~Abbasi-Yadkori, D.~P{\'a}l, and C.~Szepesv{\'a}ri, ``Improved algorithms for
  linear stochastic bandits,'' \emph{Advances in Neural Information Processing
  Systems}, pp. 2312--2320, 2011.

\bibitem{tropp2012user}
J.~A. Tropp, ``User-friendly tail bounds for sums of random matrices,''
  \emph{Foundations of computational mathematics}, vol.~12, no.~4, pp.
  389--434, 2012.

\bibitem{faradonbeh2018stabilization}
M.~K.~S. Faradonbeh, A.~Tewari, and G.~Michailidis, ``Finite time adaptive
stabilization of linear systems,'' \emph{IEEE Transactions on Automatic
	Control}, vol.~64, no.~8, pp. 3498--3505, 2019.

\bibitem{green1985persistence}
M.~Green and J.~B. Moore, ``Persistence of excitation in linear systems,'' in
  \emph{American Control Conference, 1985}.\hskip 1em plus 0.5em minus
  0.4em\relax IEEE, 1985, pp. 412--417.

\bibitem{fazel2018global}
M.~Fazel, R.~Ge, S.~Kakade, and M.~Mesbahi, ``Global convergence of policy
  gradient methods for the linear quadratic regulator,'' in \emph{International
  Conference on Machine Learning}, 2018, pp. 1466--1475.

\bibitem{abeille2018improved}
M.~Abeille and A.~Lazaric, ``Improved regret bounds for thompson sampling in
  linear quadratic control problems,'' in \emph{International Conference on
  Machine Learning}, 2018, pp. 1--9.

\bibitem{jin2018q}
C.~Jin, Z.~Allen-Zhu, S.~Bubeck, and M.~I. Jordan, ``Is q-learning provably
  efficient?'' \emph{arXiv preprint arXiv:1807.03765}, 2018.

\bibitem{tewari2008optimistic}
A.~Tewari and P.~L. Bartlett, ``Optimistic linear programming gives logarithmic
  regret for irreducible mdps,'' in \emph{Advances in Neural Information
  Processing Systems}, 2008, pp. 1505--1512.

\bibitem{bastani2017mostly}
H.~Bastani, M.~Bayati, and K.~Khosravi, ``Mostly exploration-free algorithms
  for contextual bandits,'' \emph{arXiv preprint arXiv:1704.09011}, 2017.

\bibitem{shalit2012online}
U.~Shalit, D.~Weinshall, and G.~Chechik, ``Online learning in the embedded
  manifold of low-rank matrices,'' \emph{Journal of Machine Learning Research},
  vol.~13, no. Feb, pp. 429--458, 2012.

\bibitem{polderman1986necessity}
J.~W. Polderman, ``On the necessity of identifying the true parameter in
  adaptive {LQ} control,'' \emph{Systems \& control letters}, vol.~8, no.~2,
  pp. 87--91, 1986.

\bibitem{polderman1986note}
------, ``A note on the structure of two subsets of the parameter space in
  adaptive control problems,'' \emph{Systems \& control letters}, vol.~7,
  no.~1, pp. 25--34, 1986.

\end{thebibliography}
\end{document}